\documentclass[reprint,superscriptaddress,nofootinbib,amsmath,amssymb,aps,pra]{revtex4-2}

\usepackage{graphicx}
\usepackage{dcolumn}
\usepackage{bm}
\usepackage{hyperref}
\usepackage{braket}
\usepackage{float}
\usepackage{subfig}

\begin{document}

\title{\textbf{Limits of optimal decoding under synaptic coarse-tuning}}

\author{Ori Hendler}
  \email{hendlero@post.bgu.ac.il}
  \affiliation{Department of Physics, Ben-Gurion University of the Negev, Beer-Sheva, Israel}
  \affiliation{School of Brain Sciences and Cognition, Ben-Gurion University of the Negev, Beer-Sheva, Israel}

\author{Ronen Segev}
  \affiliation{School of Brain Sciences and Cognition, Ben-Gurion University of the Negev, Beer-Sheva, Israel}
  \affiliation{Department of Life Sciences, Ben-Gurion University of the Negev, Beer-Sheva, Israel}
  \affiliation{Department of Biomedical Engineering, Ben-Gurion University of the Negev, Beer-Sheva, Israel}

\author{Maoz Shamir}
  \affiliation{Department of Physics, Ben-Gurion University of the Negev, Beer-Sheva, Israel}
  \affiliation{School of Brain Sciences and Cognition, Ben-Gurion University of the Negev, Beer-Sheva, Israel}
  \affiliation{Department of Physiology and Cell Biology, Ben-Gurion University of the Negev, Beer-Sheva, Israel}

\begin{abstract}
Sensory information propagates through successive processing stages in the brain, where synaptic weight patterns between stations determine how downstream neurons decode information from upstream populations. Although optimized synaptic connectivity can enhance information transmission, it requires precise weight tuning. Recent evidence depicting substantial synaptic volatility raises two fundamental questions: How does coarse-tuning of synaptic connectivity affect information transmission? What strategies could the nervous system employ to maintain reliable communication despite synaptic fluctuations? We addressed these questions by analyzing the signal-to-noise ratio ($SNR$) for binary stimulus discrimination under two decoding schemes: a na\"{i}ve population average and an optimized linear decoder. For the na\"{i}ve decoder, we found that $SNR$ remains largely insensitive to synaptic imprecision, since performance is already limited by correlated noise in neuronal responses. For the optimal decoder, we identified three distinct regimes, that is, weak, moderate and strong coarse-tuning. Under weak coarse-tuning, $SNR^2$ scales linearly with population size $N$. Under moderate coarse-tuning, scaling becomes sublinear. Strikingly, under strong coarse-tuning, the regime most consistent with observed neuronal heterogeneity, $SNR$ saturates and can not be improved by recruiting larger populations. This limitation persists even when incorporating feedforward or recurrent network architectures. These findings suggest that in the biologically relevant regime of strong coarse-tuning, na\"{i}ve and optimal decoders can achieve qualitatively similar performance. The analysis shows that effective readout under synaptic volatility is constrained to an invariant low-dimensional manifold aligned with the na\"{i}ve decoder, potentially pointing to a fundamental principle for robust neural computation in the face of ongoing synaptic remodeling.
\end{abstract}
\maketitle
\clearpage
\section{\label{sec:Introduction}Introduction}
One of the main enigmas in neuroscience is how sensory information is transmitted and decoded as it propagates downstream \cite{shadlen1994noise}. The profile of synaptic weights largely determines the fidelity of information transfer along the processing pathway \cite{abbott1994decoding,salinas1995transfer,seung1993simple}. Accordingly, the synaptic-weight profile can be regarded as a decoder or a readout algorithm \cite{georgopoulos1986neuronal,salinas1994vector}. One such possible decoder is a na$\ddot{\i}$ve readout that assigns uniform weights to all inputs. 
While easy to implement, it generally results in suboptimal performance \cite{zohary1994correlated,shadlen1996computational,abbott1999effect,sompolinsky2001population,hendler2025noise}. By contrast, an optimal decoder can leverage detailed knowledge of the population response statistics to achieve superior performance by fine-tuning its synaptic weights to the detailed statistics of the network activity \cite{shamir2006implications, averbeck2006effects, shamir2014emerging,snippe1992information}. The caveat is that synaptic weights undergo substantial remodeling.  

A growing body of empirical evidence has demonstrated that synapses are highly volatile \cite{chen2012clustered,cane2014relationship}. In-vivo studies show that dendritic spines, which reflect synaptic strength, exhibit substantial size fluctuations over time scales ranging from a few hours to days~\cite{loewenstein2011multiplicative, Pfeiffer2018HighTurnover, Zuo2005LongTermStability,trachtenberg2002long}. In-vitro studies of cultured cortical neurons show that synaptic sizes fluctuate dramatically at a time scale of hours~\cite{minerbi2009long, HazanZiv2020ActivityIndependent, Ren2022PredictableFluctuations,matz2010rapid}, with considerable fluctuations on a similar scale emerging even in the absence of neuronal activity~\cite{ziv2018synaptic, yasumatsu2008principles, hazan2020activity}.  

Theoretical studies typically partition synaptic weights into structured and unstructured components~\cite{kadmon2020predictive, timcheck2022optimal, shomar2017cooperative, statman2014synaptic,  shimizu2021computational}. The structured component captures the tuned connectivity, whereas the unstructured component represents random variability, which we refer to as the coarse-tuning of the weights, which may or may not fluctuate over time \cite{Hopfield1982PNAS, Amit1989MBF, susman2019stable, AmitGutfreundSompolinsky1985PRA, SompolinskyCrisantiSommers1988PRL, barral2016synaptic,loewenstein2011multiplicative, teramae2014computational}. 

Considerable theoretical attention has been devoted to querying how the brain can maintain functionality in the face of considerable coarse-tuning of synaptic weights. To date, however, the computational implications of the coarse-tuning of synaptic weights on the fidelity of information transfer in the brain remain largely unknown.

Here, we investigated the effect of the coarse-tuning of synaptic weights on readout accuracy in the framework of a modeling study. We begin by defining a statistical model of neuronal responses to the stimulus. We then define two decoders: a na$\ddot{\i}$ve decoder and an optimal one. We then apply coarse-tuning to the synaptic weights of both decoders, and investigate its effect on the decoding accuracy. Next, we consider two extensions of our readout model: a multilayer Feedforward Neural Network architecture and a Recurrent Neural Network. We show that neither can mitigate the drastic detrimental effect of coarse-tuning. The discussion centers on the implications of these findings for the theory of population coding in the brain. 

\section{Statistical Model of the neural response}
\label{sec:statistical_model_of_the_neural_response}

\begin{table}[h!]
\caption{\label{tab:glossary_alpha}Glossary of mathematical symbols used throughout the manuscript.}
\begin{ruledtabular}
\begin{tabular}{ll}
Symbol & Meaning \\ \hline
$a$ & Single neuron variance \\
$c$ & Pairwise correlation coefficient \\
$\Delta$ & Disorder in random fields \\
$g_i$ & Response selectivity of the i'th neuron\\
$\gamma$ & Coarse-tuning scaling \\
$h$ & Field \\
$J$ & Disorder in recurrent connections \\
$J_0$ & Ferromagnetic bias \\
$\kappa$ & Coarse-tuning magnitude \\
$\mu_d$ & Quenched mean response to distractor \\
$\mu_g$ & Quenched mean of response selectivity \\
$\mu_t$ & Quenched mean response to target \\
$N$ & Number of neurons \\
$s$ & Sign of the field \\
$\sigma_g^{2}$ & Quenched variance of response selectivity \\
$T$ & Temperature \\
$w_i$ & Weight of the i'th neuron\\
$\chi$ & Susceptibility 
\end{tabular}
\end{ruledtabular}
\end{table}

We model a population of $N$ neurons coding for a binary stimuli $\alpha\in\{t,d\}$ (target or distractor). We assume that the neural responses  $\mathbf{r}$, conditioned on stimulus $\alpha$, is distributed according to multivariate Gaussian statistics with mean 
\begin{equation}
    \langle r_i \rangle_{r|s} = \mu_i^\alpha,
    \label{eq:mean_firing_rate}
\end{equation}
and covariance matrix
\begin{equation}
    C_{ij}
    = a\big((1-c)\delta_{ij}+c\big),
    \label{eq:covariance_matrix}
\end{equation}
where $\langle \cdot \rangle$ denotes averaging over neural responses for a given stimulus. The parameter $a$ is the single neuron variance and $c$ is the pairwise correlation coefficient. A Gaussian approximation offers a tractable and empirically-based description of neural responses \cite{roxin2011distribution}. The population is heterogeneous, where each neuron has distinct mean response parameter $\mu_i^\alpha$ randomly drawn when the network is created and fixed thereafter. Trial-to-trial fluctuations thus arise from the noise structure whereas variability in $\mu_i^\alpha$ reflects frozen heterogeneity across neurons. Different network realizations yield different parameter sets $\boldsymbol{\mu}^\alpha$, with first-order statistics
\begin{equation}
   \ll \mu_i^t \gg = \mu_t, \quad \ll \mu_i^d \gg = \mu_d, \quad \ll g_i \gg = \mu_g,
\end{equation}
where double brackets, $\ll \cdot \gg$, denote ensemble average over realizations. The second-order statistics are
\begin{equation}
    \mathrm{Var}(g_i)=\sigma_g^2,
    \label{eq:mean_selectivity_stats}
\end{equation}
where we assume that $\mu_i^t$ and $\mu_i^d$ are uncorrelated and have equal variance $\frac{\sigma_g^2}{2}$.

We further assume that the neuronal responses exhibit stimulus selectivity, where on average, responses to the target exceed responses to the distractor. We quantify response selectivity as
\begin{equation}
    g_i = \mu_i^t - \mu_i^d.
    \label{eq:response_ selectivity}
\end{equation}

\section{The Readout Model}
\label{sec:Readout_Rule}

To examine the accuracy at which information about a stimulus can propagate downstream, we study readout performance in the framework of a two-interval-two-alternative forced-choice task. In this paradigm, one interval presents the target stimulus and the other the distractor, with their order randomized across trials. Without loss of generality, we assume that the target is presented in the first interval for all analytical calculations. The linear readout discriminates between the two alternatives according to the sign of the field:
\begin{equation}
    h \equiv \mathbf{w}^{\top}\big(\mathbf{r}^t-\mathbf{r}^d\big).
\label{eq:h_def}
\end{equation}
Thus, the linear readout is fully defined by its weight vector $\mathbf{w}$. The binary decision follows from the sign of this field:
\begin{equation}
   s=\mathrm{sign}(h),
   \label{eq:decision_variable}
\end{equation}
where $s=+1$ indicates "target first" (correct under our assumption) and $s=-1$ indicates  "target second" (incorrect).
In our Gaussian response model, the field \(h\) varies from trial to trial with mean \(\langle h\rangle=\mathbf{w}^{\top}\big(\boldsymbol{\mu}^{t}-\boldsymbol{\mu}^{d}\big)\) and variance \(\langle(\delta h)^2\rangle=2\,\mathbf{w}^{\top}\mathbf{C}\mathbf{w}\), contingent on the specific realization of $\mathbf{w}$. For a given realization, the probability of error is therefore 
\begin{equation}
    P_{\text{err}} = \int_{SNR}^{\infty} \frac{1}{\sqrt{2\pi}} e^{-u^2/2} \, du,
    \label{eq:error_probability}
\end{equation}
where the signal-to-noise ratio ($SNR$) squared is given by
\begin{equation}
\begin{aligned}
SNR^2&=\frac{(\mathbf{w}^{\top}  \mathbf{g}\big)^2}{2\mathbf{w}^{\top}\mathbf{C}\mathbf{w}}, & signal&={\langle h\rangle}, & noise&=\sqrt{\langle(\delta h)^2\rangle}.
\end{aligned}
\label{eq:SNR_def}
\end{equation}
The probability of error is fully determined by the $SNR$. For large values of the $SNR$, a standard asymptotic expansion yields
\begin{equation}
    P_{\mathrm{err}}=\frac{e^{-SNR^{2}/2}}{\sqrt{2\pi}\,SNR}\left( 1-\mathcal{O}(SNR^{-2}) \right).
    \label{eq:asymptotic_Pe}
\end{equation}
where the error probability decays exponentially with $SNR^{2}$ up to algebraic corrections. Accordingly, we study how different choices of the weight vector $\mathbf{w}$ impact the $SNR$ and hence task performance, see Fig.\ref{fig:fig1a}.

\section{Readout Performance}
\label{sec:readout_performance}

\subsection{Na\"{i}ve Decoder Performance}
We analyze the $SNR$ as a function of  different choices of readout weights, starting with the Na\"{i}ve decoder.
\begin{figure}[h!]
    \subfloat{%
        \begin{tabular}[b]{@{}l@{}}
        \hspace{-5mm}\textbf{(a)}\\[-1mm]
        \includegraphics[width=0.32\textwidth]{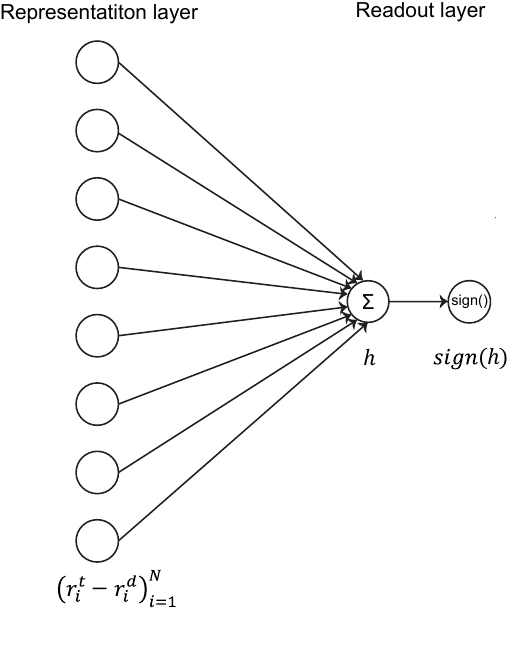}
        \end{tabular}%
        \label{fig:fig1a}}%
    \hspace{0.8em}
    \subfloat{%
        \begin{tabular}[b]{@{}l@{}}
        \hspace{-5mm}\textbf{(b)}\\[-4mm]
        \includegraphics[width=0.32\textwidth]{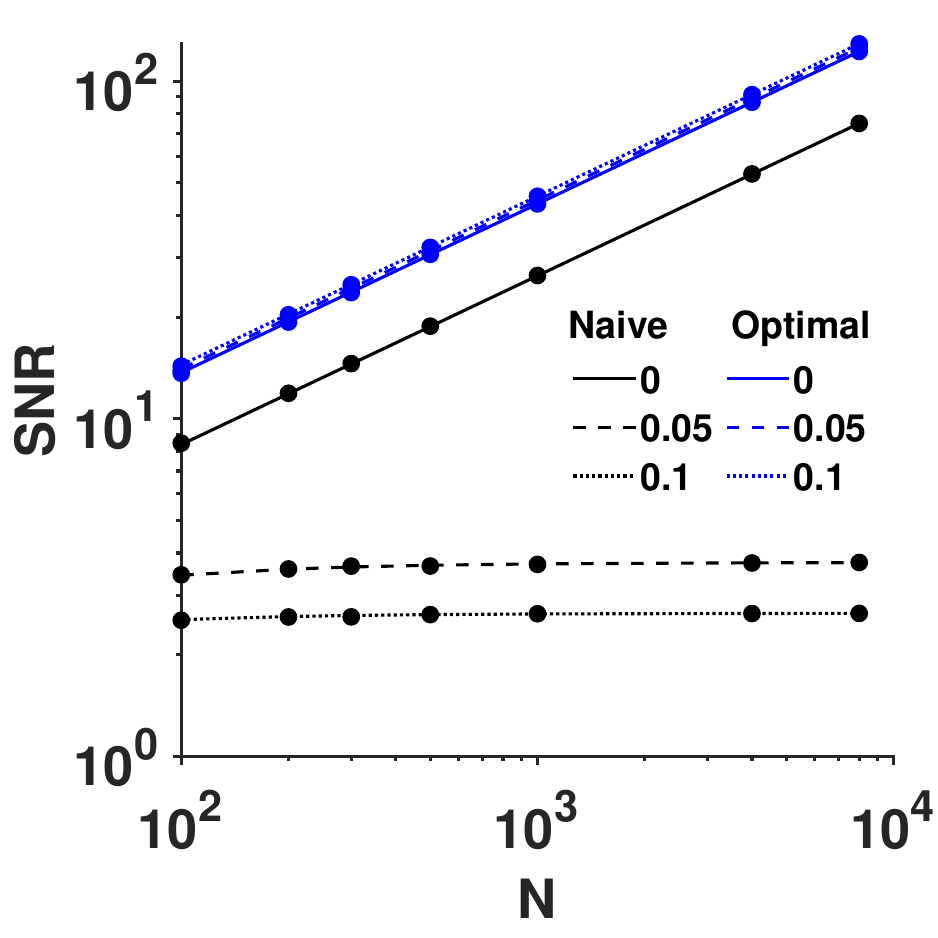}
        \end{tabular}%
        \label{fig:fig1b}}%
\captionsetup{justification=raggedright,singlelinecheck=false}
\caption{\textbf{Readout model of the na\"ive and optimal decoders.}
\textbf{(a)} Schematic illustration of the model architecture.
\textbf{(b)} The signal-to-noise ratio ($SNR$) is plotted as a function of population size $N$ for the optimal (blue) and na\"ive (black) decoders.
Different values of the pairwise correlation coefficient $c=0, 0.05, 0.1$ are depicted by solid, dashed, and dotted lines, respectively.}
\label{fig:fig1}
\end{figure}
Since the response to the target is larger on average than the response to the distractor, the na\"{i}ve approach consists of assigning equal weights to all neurons:
\begin{equation}
    w_i^{\text{na\"{i}ve}}=\frac{1}{N}.
    \label{eq:naive_weights}
\end{equation}
The $noise^2$ (see Eq.\eqref{eq:SNR_def}) of the na\"{i}ve decoder is $ \mathcal{O}( N^0 )$ and does not depend on the specific realization of quenched disorder, see Appendix~\ref{app:Naive_SNR2}. The $signal^2$ of the na\"{i}ve decoder is a random variable that depends on the specific realization of the system; i.e., the choice of $ \{ g_i \}_{i=1}^N $. The $signal^2$ has a mean of $ \mathcal{O}( N^0)$ and variance $\mathcal{O}( N^{-1})$ across different realizations. Thus, for large populations (\(N \to \infty\)), the $signal^2$ of a typical realization is equal to its quenched mean, up to fluctuations that vanish in the limit of large $N$. This property of a random variable whose fluctuations vanish relative to its mean is referred to as self-averaging. In the limit of large populations, a self-averaging variable can be replaced by its quenched mean. The quenched mean of the squared signal‐to‐noise ratio for the na$\ddot{\i}$ve decoder is (see Appendix~\ref{app:Naive_SNR2}):
\begin{equation}
    \ll SNR^2(\mathbf{w}^{\text{na\"{i}ve}})\gg\approx
\begin{cases}
    \frac{N\mu_g^2}{2a} & \text{, if } c=0\\
    \frac{N\mu_g^2}{2a}\left( \frac{1}{(1-c)+\frac{N}{N_{\mathrm{eff}}}} \right)& \text{, if } c>0
\end{cases}
\label{eq:SNR_naive_Neff}
\end{equation}
where we define \(N_{\mathrm{eff}} \equiv \frac{1}{c}\). When response fluctuations of different neurons are uncorrelated, $c=0$, the $SNR^2$ scales linearly with the population size, $N$, $\ll SNR^2(\mathbf{w}^{\text{na\"{i}ve}}) \gg \propto N I_0$, where $I_0 = \frac{\mu_g^2}{2a}$ is the (quenched mean of the) $SNR^2$ of the single neuron, Fig.\ref{fig:fig1b} (solid black line). For $c>0$, the $SNR^2$ scales linearly with the population size, $\ll SNR^2 (\mathbf{w}^{\text{na\"{i}ve}}) \gg \propto N I_0$, for small populations, $N \ll N_{eff}$. In the limit of large populations, $N \gg N_{eff}$, the $SNR^2$ of the na\"{i}ve readout saturates to $\ll SNR^2(\mathbf{w}^{\text{na\"{i}ve}}) \gg\propto N_{eff} I_0$, Fig.\ref{fig:fig1b}. Note that due to self-averaging, the error bars denoting the standard deviation of the $SNR^2$ across different realizations of the quenched disorder are barely visible.

\subsection{Optimal Decoder Performance}
The optimal weights that maximize the probability of correct discrimination are given by:
\begin{equation}
    \mathbf{w}^{\text{opt}} = \mathbf{C}^{-1} \mathbf{g}.
    \label{eq:optimal_weights}
\end{equation}
In our model; i.e., Eq.~\ref{eq:covariance_matrix}, this yields:
\begin{equation}
    w_i^{\mathrm{opt}} = \frac{1}{N a(1 - c)} \left( g_i - \frac{N c}{(1 - c) + N c} \bar{g} \right),
    \label{eq:opt_weights_explicit}
\end{equation}
where \(\bar{g} = \frac{1}{N} \sum_{j=1}^N g_j\) is the population-average response selectivity. In the limit of a large population (\(N \to \infty\))
the weights are given by
\begin{equation}
    w_i^{\mathrm{opt}} \approx \frac{1}{N a(1 - c)} \left( g_i - \bar{g} \right).
    \label{eq:opt_large_N}
\end{equation}
Hence, the optimal decoder uses the heterogeneity to extract information from a subspace that is orthogonal to the space with large noise correlations. We use this approximation, Eq.~\ref{eq:opt_large_N}, for the analysis below. 
The $SNR^2$ of the optimal decoder (Eq.~\eqref{eq:opt_weights_explicit}) is given by (see Appendix~\ref{app:Optimal_SNR2}):
\begin{equation}
    \ll SNR^2(\mathbf{w}^{\mathrm{opt}})\gg 
    = 
    \frac{
    \bigl(1+(N-2)c\bigr)\sigma_g^2 +
    \bigl(1-c\bigr)\mu_g^2
    }{2a(1-c)\Bigl(\frac{1+c(N-1)}{N}\Bigr)}.
    \label{eq:SNR_opt_final}
\end{equation}
In a homogeneous population; i.e., $\sigma_g^2=0$, the optimal readout is the na\"ive readout and its $SNR$ is limited by the noise correlations. In a heterogeneous population, \(\sigma_g^2 \neq 0\), the $SNR$ of the optimal readout is superior to that of the na\"ive one. In particular its $SNR^2$ scales linearly with the population size, the leading term:
\(
    \ll SNR^2(\mathbf{w}^{\text{opt}}) \gg \propto \frac{N\sigma_g^2}{2a(1-c)}
\) (for $c\neq0$), see Fig.\ref{fig:fig1b} (blue lines).
However, these results rely on the ability of the system to finely tune the readout weights to the specific realization of the response heterogeneity; e.g.,  Eq.~\ref{eq:opt_weights_explicit}.

\section{Coarse-tuned readout model}
\label{sec:coarse_tuning}
Up to now, the quenched variability was only applied to the neuronal responses (i.e., $ \{ g_i \}_{i=1}^N$), but not to the synaptic weights. To model synaptic volatility we introduce additive quenched fluctuations to the weights,
\begin{equation}
    \tilde{w}_i = \frac{w_i}{\sqrt{N} \|\mathbf{w}\|} + \xi_i, \quad
    \xi_i \stackrel{\text{i.i.d.}}{\sim} \mathcal{N}\left(0,\kappa^2 N^{\gamma-1}\right),
    \label{eq:perturbed_weights}
\end{equation}
where the fluctuations $\xi_i$ are independent and identically distributed Gaussian variables, with zero mean and variance \( \kappa^2 N^{\gamma-1}\). The parameter $\kappa$ governs the magnitude of the fluctuation; i.e., the coarse-tuning magnitude. The parameter $\gamma$ determines the scaling of the fluctuation variance with the population size, $N$; i.e., the coarse-tuning scaling. 

Thus, the weight $\tilde{w}_i$ is expressed as the sum of a structured component of order $\mathcal{O}(N^{-1})$ and an unstructured component that scales like $\mathcal{O}(N^\gamma)$. This decomposition into structured and unstructured components is a classic feature of  theoretical models, such as the Hopfield model \citep{Hopfield1982PNAS}, the Sherrington–Kirkpatrick (SK) model \citep{sherrington1975solvable}, and spin-glass models of neural networks \citep{AmitGutfreundSompolinsky1985PRL, AmitGutfreundSompolinsky1985PRA}, as well as more recent works \citep{kadmon2020predictive,timcheck2022optimal,KadmonSompolinsky2015PRX,AljadeffSternSharpee2015PRL}. In these models, the structured component of the weight vector $\mathbf{w}$ typically scales as $\mathcal{O}(N^{-1})$, whereas the unstructured component scales as $\mathcal{O}(N^{-\frac{1}{2}})$, which in our notation corresponds to $\gamma = 0$.

\section{Impact of Coarse-Tuning on Readout Performance}
\label{Impact_of_Coarse_Tuning_on_Readout_Performance}

\begin{figure*}[t]  
\centering
\includegraphics[width=\textwidth]{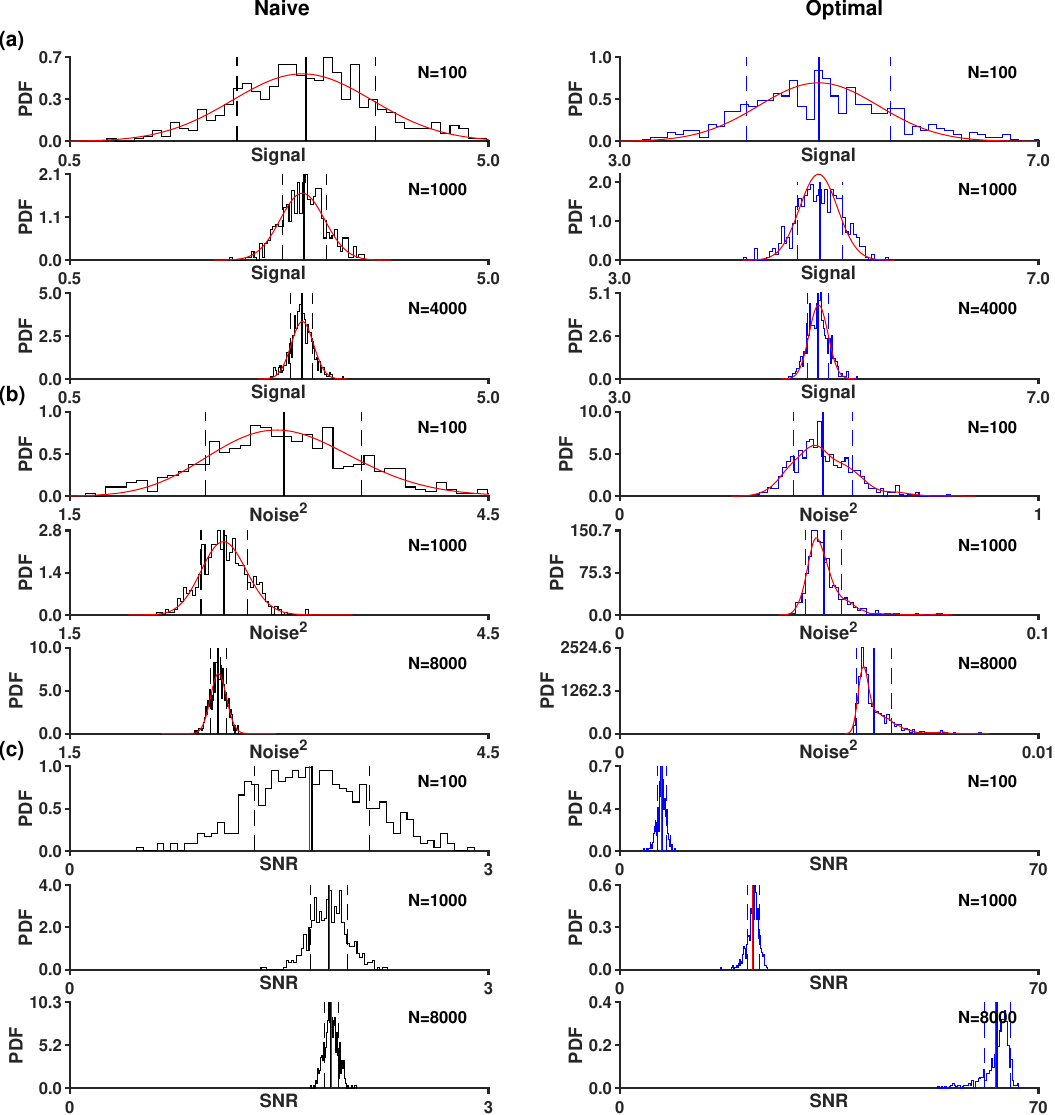}
\captionsetup{justification=raggedright,singlelinecheck=false}
\caption{\label{fig:fig2}
\textbf{Histograms of signal, noise}$\bm{^2}$\textbf{, and SNR.} 
Distributions of the na\"{i}ve (black) and optimal (blue) decoders across
population sizes $N\in\{100,1000,4000\}$ under coarse-tuning
with $\kappa=1$ and $\gamma=-1$.
\textbf{(a)} $signal$ distribution.
\textbf{(b)} $noise^2$ distribution.
\textbf{(c)} $SNR$ distribution.
Solid vertical lines indicate the mean; dashed lines indicate the  $\pm1$ standard deviation.
Red curves show analytical predictions derived in the main text.
}
\end{figure*}

The analysis of  the effects of coarse-tuning on the $SNR$ of the na\"ive and optimal readouts (below) identifies three regimes: weak ($\gamma=-1$), moderate ($-1<\gamma<0$) and strong ($\gamma=0$) coarse-tuning. For weak coarse-tuning ($\gamma=-1$), the $signal$ is self-averaging for both readouts, with mean $ \mathcal{O}(N^0)$ and variance of order $\mathcal{O}(N^{-1})$, see Appendix~\ref{app:Naive_Decoder_under_Coarse_Tuning},~\ref{app:Optimal_Decoder_under_Coarse_Tuning}. Fig.\ref{fig:fig2}a presents the quenched distribution of the $signal$ for weak coarse-tuning across the different realizations of response selectivity, \( \{ g_i \}_{i=1}^N \), and weights, $\{\tilde{w}_i\}_{i=1}^N$, where each row corresponds to a different population size, $N$. As shown in the figure, the variability of the $signal$ vanishes with increasing $N$. For weak coarse-tuning, the $noise^2$ of the na\"{i}ve decoder is self-averaging, Fig.\ref{fig:fig2}b (left column). However, the $noise^2$ of the optimal decoder is not self-averaging, Fig.\ref{fig:fig2}b (right column), with mean $\mathcal{O}(N^{-1})$ and variance $\mathcal{O}(N^{-2})$. Nevertheless, for both decoders, the $SNR$ is well-approximated by $\ll SNR \gg \approx \frac{\ll signal \gg}{\sqrt{\ll noise^2 \gg}}$, Fig.\ref{fig:fig2}c. 

Similar behavior is found for moderate coarse-tuning. Specifically, the $signal^2$ of both readouts is self-averaging. The  $noise^2$ is self-averaging for the na\"{i}ve decoder (Appendix~\ref{app:Naive_Decoder_under_Coarse_Tuning}), but not for the optimal decoder (Appendix~\ref{app:Optimal_Decoder_under_Coarse_Tuning}). In the strong coarse-tuning regime ($\gamma = 0$), neither the $signal$ nor the $noise$ are self-averaging. Nevertheless, $\ll SNR \gg$ is well-approximated by the ratio of the averaged $signal$ to the square root of the averaged $noise^2$; i.e., $ \ll SNR \gg \approx \frac{\ll signal \gg}{\sqrt{\ll noise^2 \gg}}$. The quenched mean $\ll SNR \gg$ provides a fair approximation of the typical $SNR$. We therefore use the $signal$ and $noise^2$ to quantify the $SNR$ below. 
\begin{figure*}[!htbp]
\centering
\subfloat{%
    \begin{tabular}[b]{@{}l@{}}
    \textbf{(a)}\\
    \includegraphics[width=0.31\textwidth]{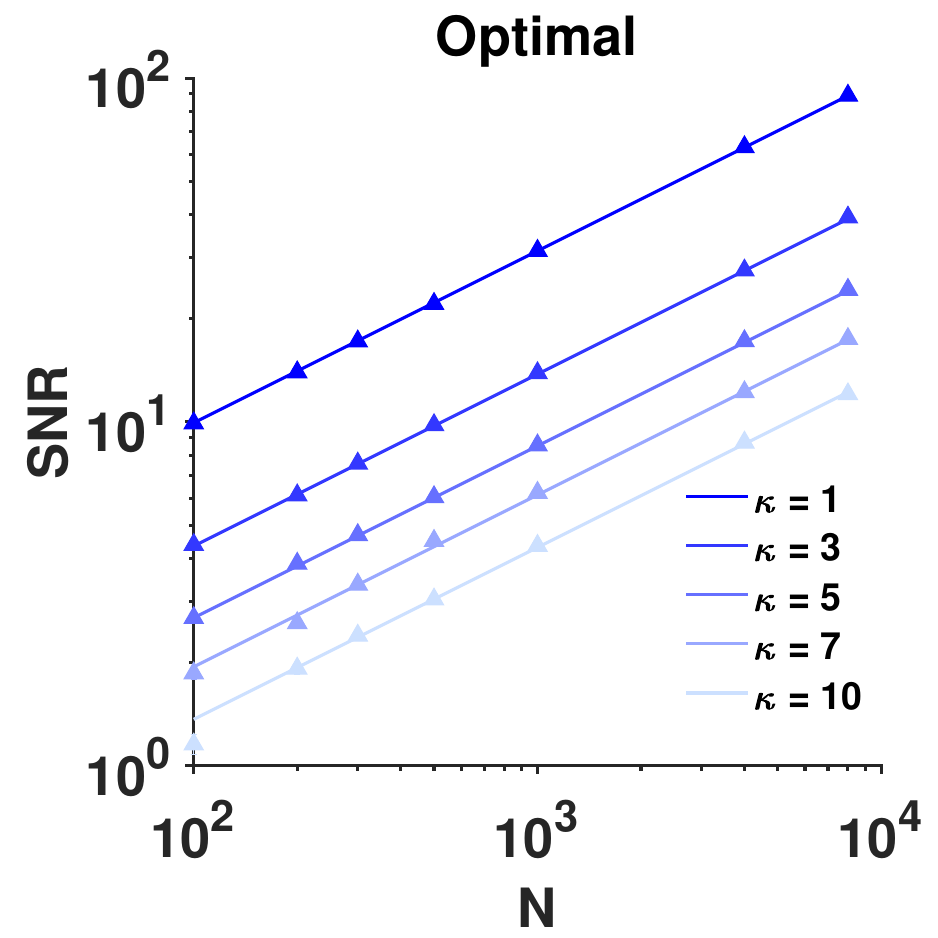}
    \end{tabular}%
    \label{fig:fig3a}}%
\hfill
\subfloat{%
    \begin{tabular}[b]{@{}l@{}}
    \textbf{(b)}\\
    \includegraphics[width=0.31\textwidth]{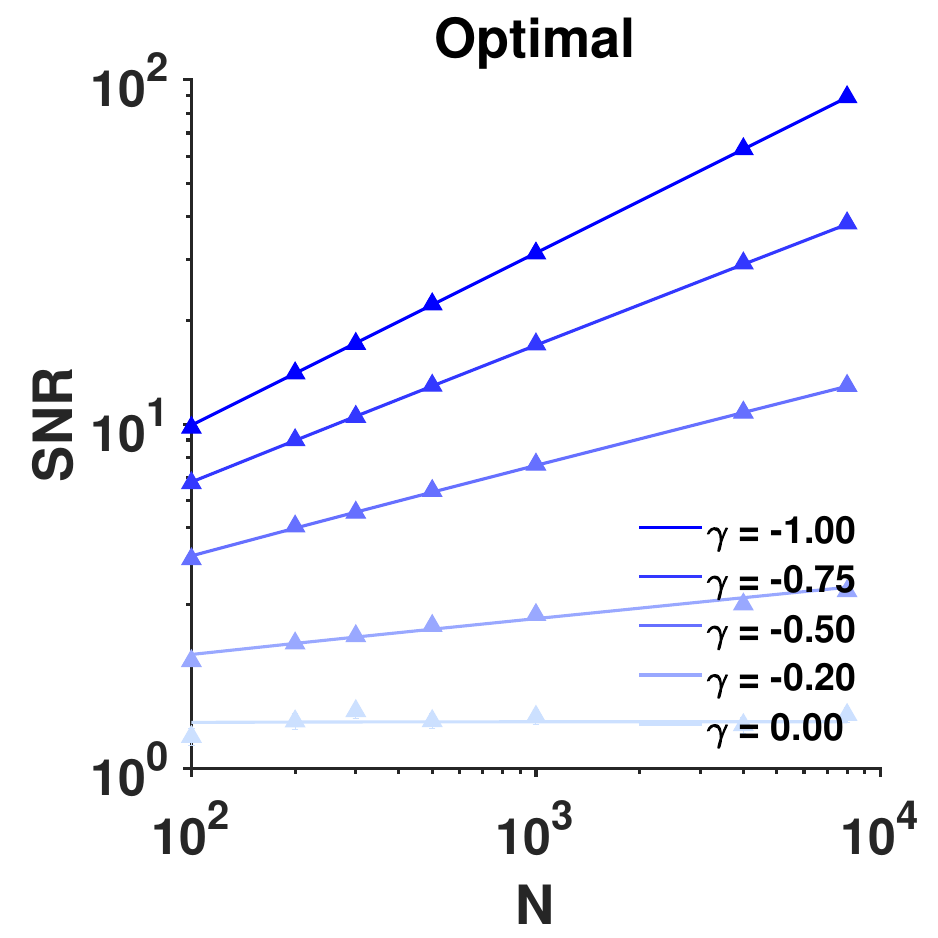}
    \end{tabular}%
    \label{fig:fig3b}}%
\hfill
\subfloat{%
    \begin{tabular}[b]{@{}l@{}}
    \textbf{(c)}\\
    \includegraphics[width=0.31\textwidth]{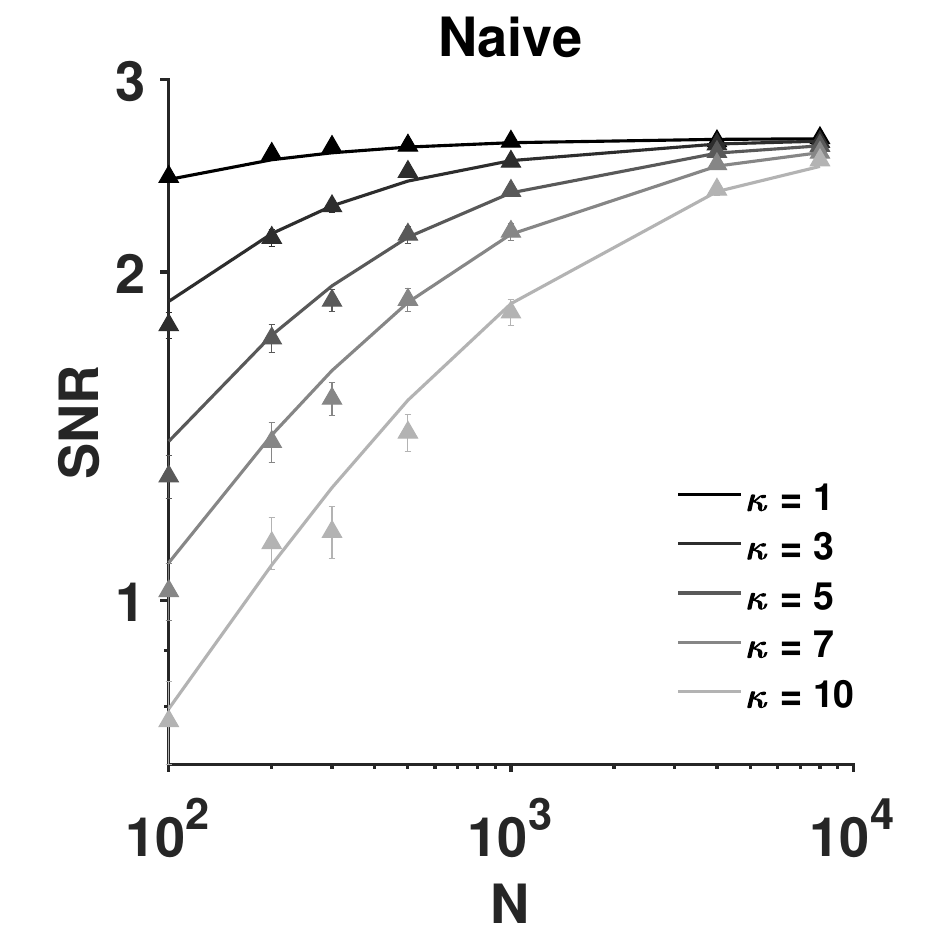}
    \end{tabular}%
    \label{fig:fig3c}}%
\par\medskip
\begin{minipage}{\textwidth}
  \centering
\subfloat{%
    \begin{tabular}[b]{@{}l@{}}
    \textbf{(d)}\\
    \includegraphics[width=0.31\textwidth]{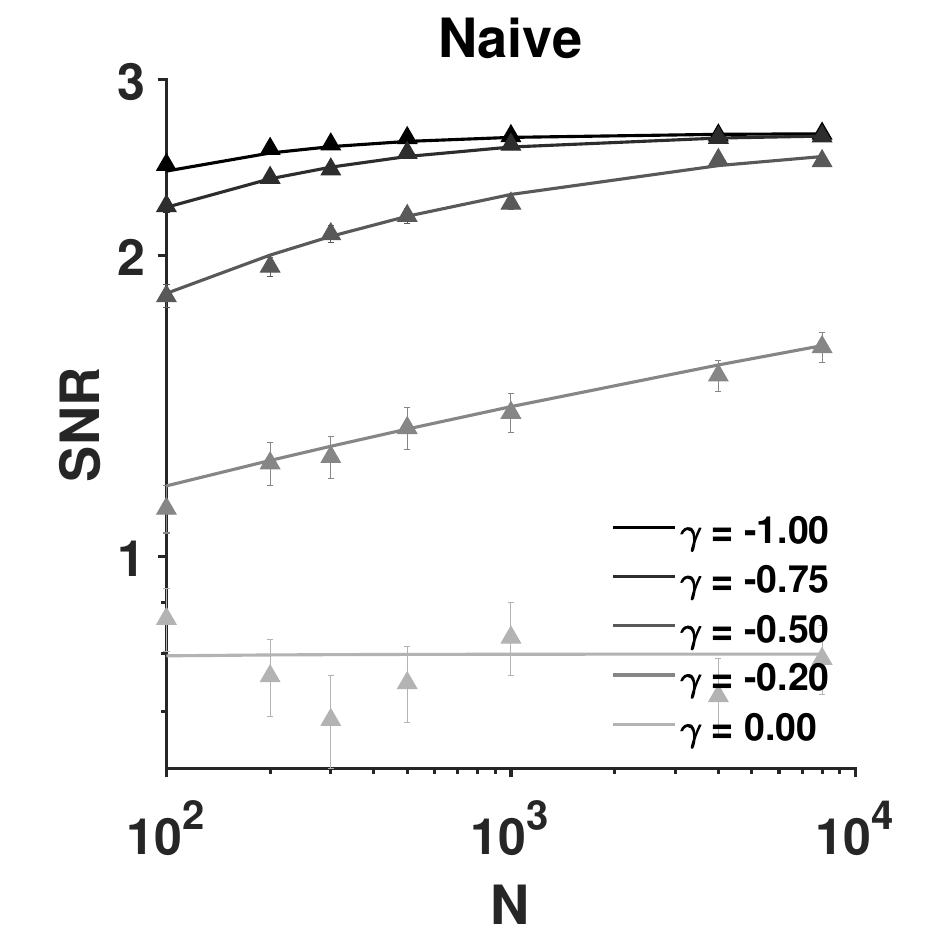}
    \end{tabular}%
    \label{fig:fig3d}}%
  \hspace{0.06\textwidth}
\subfloat{%
    \begin{tabular}[b]{@{}l@{}}
    \textbf{(e)}\\
    \includegraphics[width=0.31\textwidth]{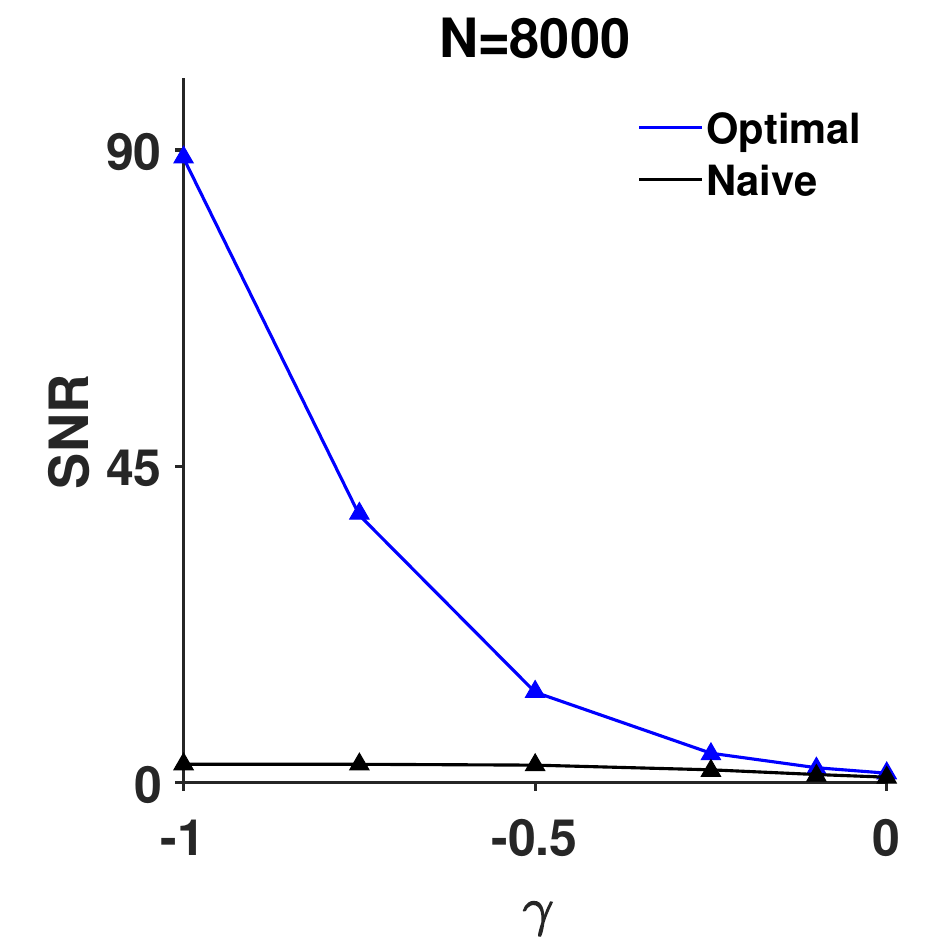}
    \end{tabular}%
    \label{fig:fig3e}}%
\end{minipage}
\captionsetup{justification=raggedright,singlelinecheck=false}
\caption{\label{fig:fig3}\textbf{$\boldsymbol{ SNR}$ scaling with population size under coarse-tuning.}
\textbf{(a)} $SNR$ versus population size $N$ for the optimal decoder, with varying magnitudes of coarse-tuning, $\kappa\in\{1, 3, 5, 7, 10\}$ at fixed $\gamma=-1$.
\textbf{(b)} $SNR$ versus $N$ for the optimal decoder, with varying scalings of coarse-tuning, $\gamma\in\{-1, -0.75, -0.5, -0.2, 0\}$ at fixed $\kappa=1$.
\textbf{(c,d)} Same as (a,b), but for the na\"{i}ve decoder.
\textbf{(e)} $SNR$ versus $\gamma$ at fixed population size $N=8000$ and $\kappa=1$.
In all panels, the color gradient from dark to light indicates increased coarse-tuning.}
\end{figure*}

We find that for large $N$ the quenched mean of the $SNR$ of the na\"ive and of the optimal readouts is well-approximated by (see Appendix~\ref{app:Naive_Decoder_under_Coarse_Tuning},~\ref{app:Optimal_Decoder_under_Coarse_Tuning}):
\begin{equation}
    \ll SNR(\tilde{\mathbf{w}}^{\text{na\"ive}})\gg
    \approx
    \frac{
    \mu_g 
    }{
    \sqrt{2a} \sqrt{\frac{1+(N-1)c}{N} +  \kappa^2 N^\gamma }
    },
    \label{eq:SNR_naive_perturbed}
\end{equation}
\begin{equation}
    \ll SNR(\tilde{\mathbf{w}}^{\text{opt}})\gg
    \approx
    \frac{
    \sigma_g
    }{
    \sqrt{2a} \sqrt{\frac{(1-c)}{N} + \kappa^2 N^\gamma}
    }.
    \label{eq:SNR_opt_perturbed}
\end{equation}
As shown in Eqs.~\eqref{eq:SNR_naive_perturbed}-\eqref{eq:SNR_opt_perturbed}
the nominator is independent of the population size and the scaling of the denominator with $N$ depends on $\gamma$. 

In the weak coarse-tuning regime, $\gamma=-1$, the $SNR^2$ of the optimal decoder scales linearly with $N$. The slope of the linear scaling is determined by the magnitude of the quenched fluctuations, $\kappa$ (see Fig.\ref{fig:fig3a}). For moderate coarse-tuning, $-1< \gamma < 0$, the $SNR^2$  grows with the population size, albeit in a sublinear manner, scaling as $N^{-\gamma}$, Fig.\ref{fig:fig3b}. For strong coarse-tuning, $\gamma=0$, the $SNR$ saturates to an asymptotic value, $\frac{\sigma_g}{\sqrt{2a\kappa^2}}$, in the limit of large $N$, Fig.\ref{fig:fig3b} and Fig.\ref{fig:fig3e}.
   
The $SNR$ of the na\"{i}ve decoder saturates in all coarse-tuning regimes. For weak and moderate coarse-tuning, $-1 \leq \gamma <0$, the asymptotic value of the na\"ive $SNR$ is determined by the noise correlations, $ \lim_{N \rightarrow \infty} SNR = \frac{\mu_g}{\sqrt{2ac}}$, and is independent of the quenched disorder, Fig.\ref{fig:fig3c} and Fig.\ref{fig:fig3d}. In the case of strong coarse-tuning, $\gamma=0$, the $SNR$ of the na\"ive readout saturates to  $\lim_{N \rightarrow \infty} SNR = \frac{\mu_g}{\sqrt{2a(c+\kappa^2)}}$, Fig.\ref{fig:fig3e}.

In summary, in weak and moderate levels of coarse-tuning, $-1 \leq \gamma <0$, the  optimal readout is superior to that of the na\"ive readout, and its $SNR$ increases with the population size. In the strong coarse-tuning regime, $\gamma = 0$, the $SNR$ of both readouts saturate to a finite limit. Moreover, in this case, the performance of the na\"ive readout can be superior to that of the optimal readout, depending on the choice of parameters.  This raises the question of whether the performance of the optimal readout in the case of strong coarse-tuning can be improved by considering a more elaborate readout structure incorporating more layers and recurrent connectivity. 

\section{Readout Layer and Further Processing}

\subsection{Feedforward Neural Network}
\begin{figure*}[!htbp]
\begin{minipage}{\textwidth}\centering
\subfloat{%
    \begin{tabular}[b]{@{}l@{}}
    \textbf{(a)}\\
    \includegraphics[width=0.45\textwidth]{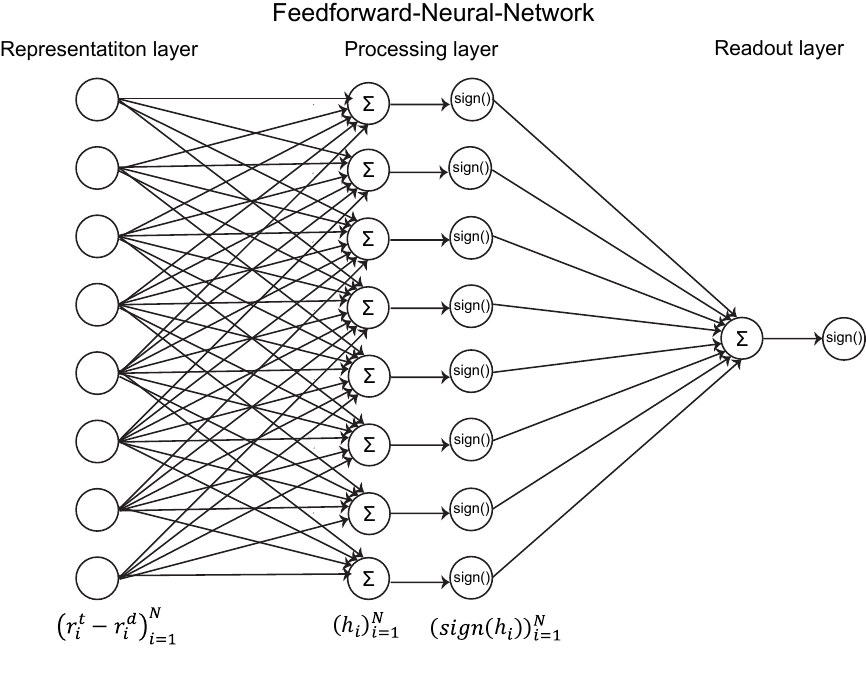}
    \end{tabular}%
    \label{fig:fig4a}}%
\hfill
\subfloat{%
    \begin{tabular}[b]{@{}l@{}}
    \textbf{(b)}\\
    \includegraphics[width=0.45\textwidth]{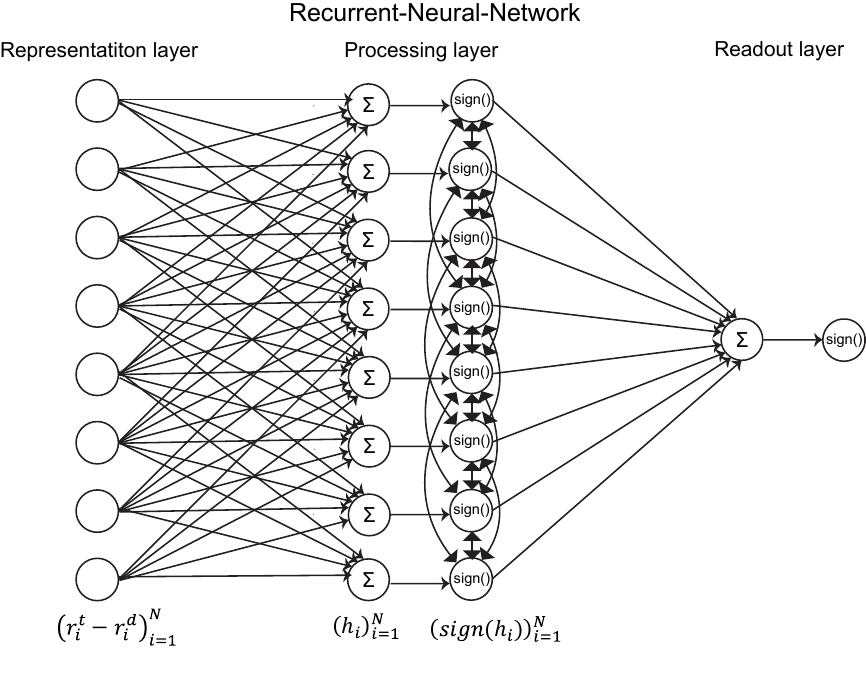}
    \end{tabular}%
    \label{fig:fig4b}}%
\end{minipage}
\par\vspace{-2em}
\begin{minipage}{\textwidth}\centering
\subfloat{%
    \begin{tabular}[b]{@{}l@{}}
    \textbf{(c)}\\
    \includegraphics[width=0.4\textwidth]{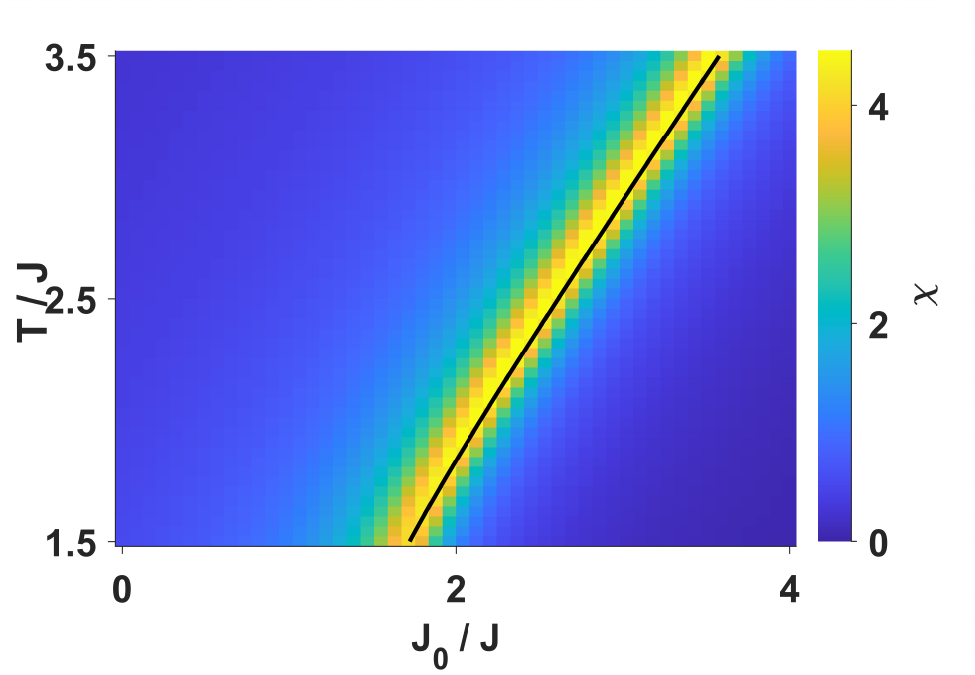}
    \end{tabular}%
    \label{fig:Heatmap_Ru_delta_half}}%
\hfill
\subfloat{%
    \begin{tabular}[b]{@{}l@{}}
    \textbf{(d)}\\
    \includegraphics[width=0.4\textwidth]{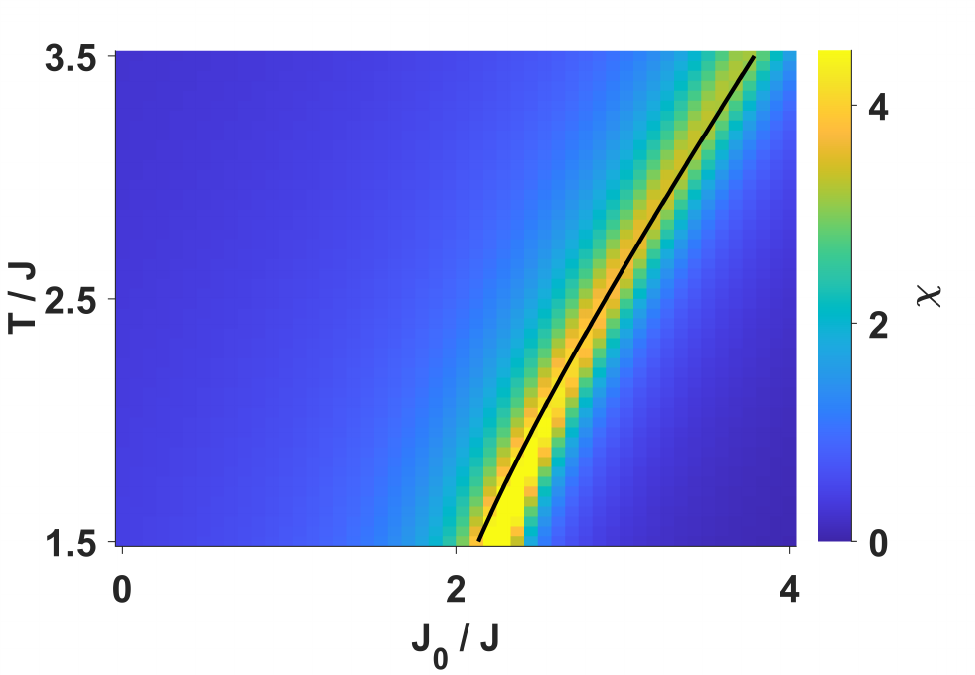}
    \end{tabular}%
    \label{fig:Heatmap_Ru_delta_one}}%
\end{minipage}
\par\vspace{-2.5em}
\begin{minipage}{\textwidth}\centering
\subfloat{%
    \begin{tabular}[b]{@{}l@{}}
    \textbf{(e)}\\
    \includegraphics[width=0.4\textwidth]{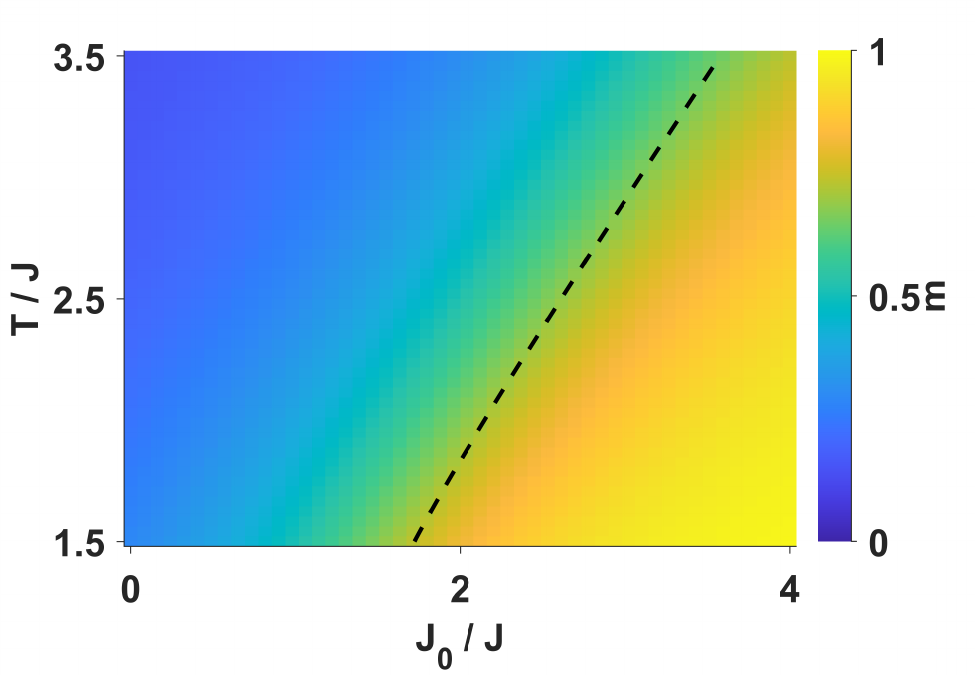}
    \end{tabular}%
    \label{fig:Heatmap_m_delta_half}}%
\hfill
\subfloat{%
    \begin{tabular}[b]{@{}l@{}}
    \textbf{(f)}\\
    \includegraphics[width=0.4\textwidth]{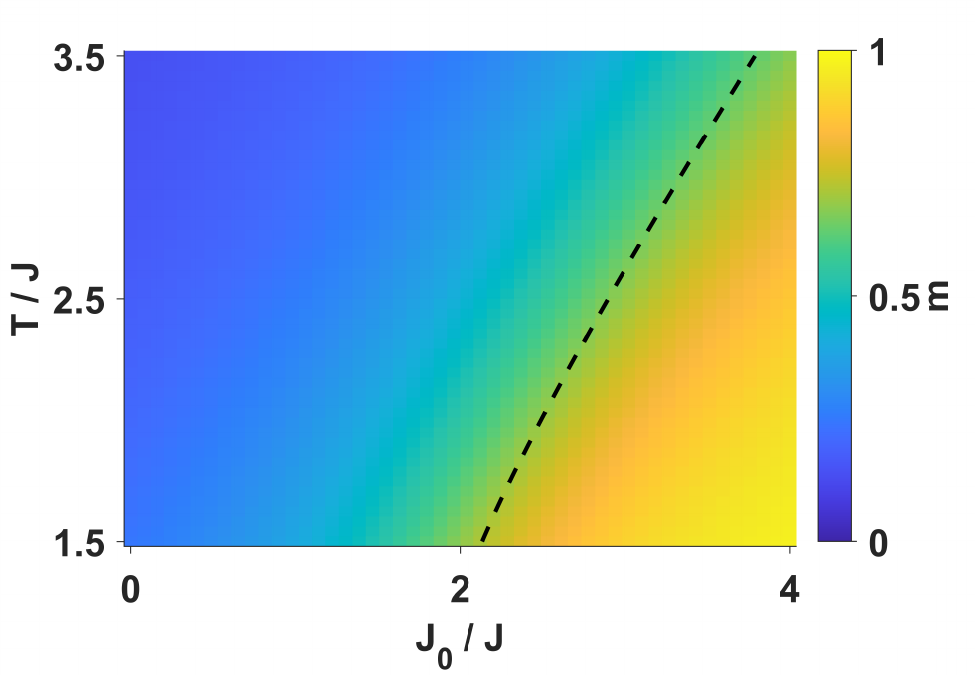}
    \end{tabular}%
    \label{fig:Heatmap_m_delta_one}}%
\end{minipage}
\par\vspace{-2.5em}
\begin{minipage}{\textwidth}\centering
\subfloat{%
    \begin{tabular}[b]{@{}l@{}}
    \textbf{(g)}\\
    \includegraphics[width=0.4\textwidth]{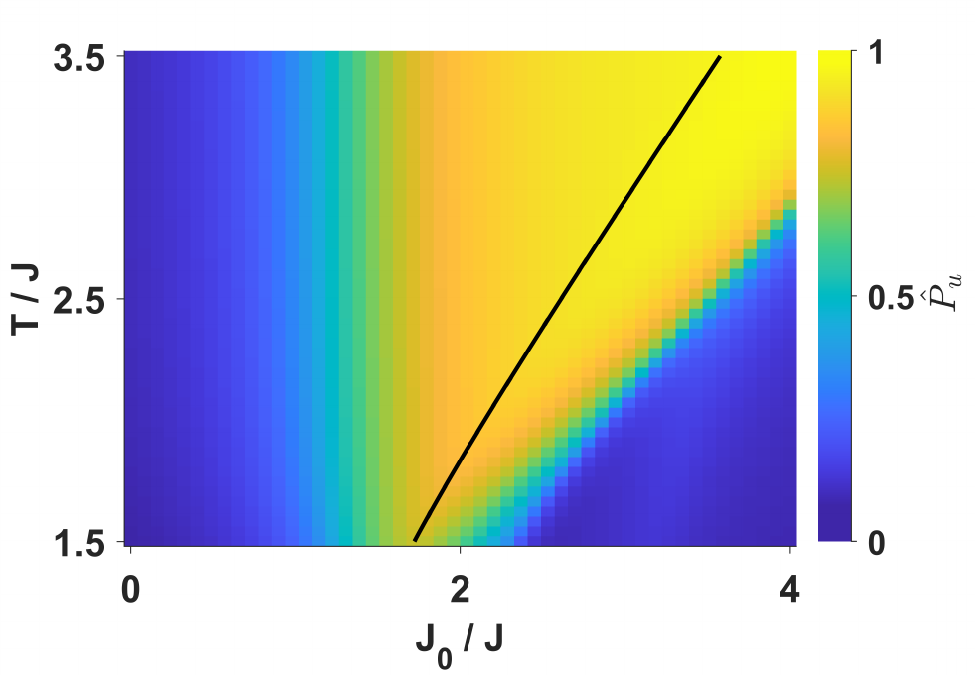}
    \end{tabular}%
    \label{fig:Heatmap_abs_cu_uh}}%
\hfill
\subfloat{%
    \begin{tabular}[b]{@{}l@{}}
    \textbf{(h)}\\
    \includegraphics[width=0.4\textwidth]{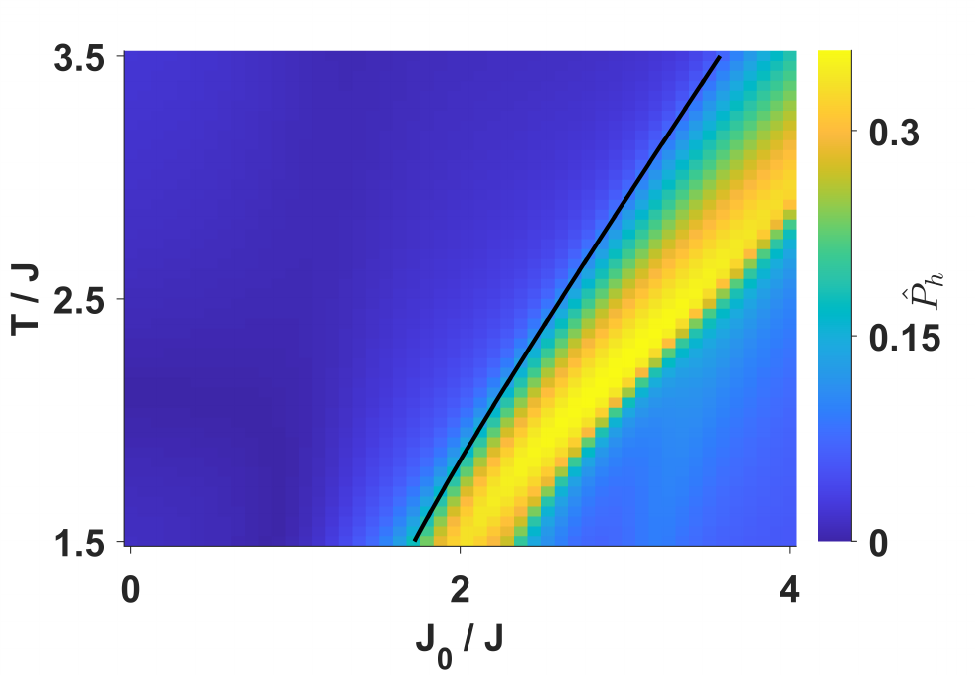}
    \end{tabular}%
    \label{fig:Heatmap_abs_ch_uh}}%
\end{minipage}
\captionsetup{justification=raggedright,singlelinecheck=false}
\caption{\label{fig:fig4}\textbf{Coarse-tuning of the Feedforward Neural Network and the Recurrent Neural Network.}
\textbf{(a,b)} Schematic illustration of the (a) Feedforward Neural Network and (b) Recurrent Neural Network architectures.
\textbf{(c,d)} Susceptibility in the $T/J$--$J_0/J$ parameter space for (c) $\Delta=0.5$ and (d) $\Delta=1$. The solid black line marks the phase transition.
\textbf{(e,f)} Magnetization heat maps for (e) $\Delta=0.5$ and (f) $\Delta=1$. The dashed black line indicates the phase transition for the corresponding $h_0=0$ case.
\textbf{(g,h)} Projection of the signal onto (g) the uniform direction and (h) the random local field.}
\end{figure*}

To incorporate an additional processing layer, we treat the binary decision variable $s$ (Eq.~\ref{eq:decision_variable})
as a single neuron in the processing layer of $N$ neurons, $ \{ s_i \}_{i-1}^N$ with $s_i = \mathrm{sign}(h_i)$ (see Fig.\ref{fig:fig4a}), where $h_i$ is the local field of neuron $i$ and is given by 
\begin{equation}
    h_i=\sum_j \tilde{w}_{i,j}(r_j^t-r_j^d),
\end{equation} 
As above, we assume that the feedforward weight matrix, $\tilde{\mathbf{w}} = \mathbf{w} + \boldsymbol{\xi}$, is a sum of two terms. The structured component, $ \mathbf{w}$, is identical for all neurons, $w_{i,j} = w_{i',j} \ ( \forall i, i')$. The second term, $\boldsymbol{\xi}$, is a random matrix representing the quenched noise of the synaptic weights. The elements $\{ \xi_{i,j} \}$ are assumed to be i.i.d. Gaussian random variables with zero mean and 
$
    \ll \xi_{i,j} \xi_{i',j'} \gg
    = \delta_{i,i'} \delta_{j,j'} \kappa^2 N^{\gamma-1}$. 

Given the synaptic weights, $ \mathbf{w}$, the local fields, $ \{ h_i \}$, are random variables with a mean over trials $\langle h_i\rangle = \sum_j \tilde{w}_{i,j}g_j$. We denote the average taken over both trials and realizations by $[\cdot]$, yielding $[h_i]=\sigma_g \equiv h_0$. The covariance, \(
    \mathrm{cov}(h_i,h_k) = [h_i h_k] - h_0^2
\), is given by (see Appendix~\ref{app:ffnn}) 
\begin{equation}
\label{eq:Cov_h}
    \mathrm{cov}(h_i,h_k) =
        \dfrac{a(1-c)}{N}+N^{\gamma}\kappa^2(\mu_g^2+\sigma_g^2+a) \delta_{ik}. 
\end{equation} 
The first term on the right-hand side of Eq.~\ref{eq:Cov_h} arises from the fact that all neurons in the processing layer receive identical structured inputs. This term represents the residual noise present in the optimal readout even under perfect fine-tuning of the weights. 

In what follows, we neglect this contribution and approximate the distribution of the random fields, 
$\{ h_i \}$, by independent Gaussian variables with
\begin{equation}
    \Pr(h_i) 
    = \frac{1}{ \sqrt{2\pi} \Delta} 
    \exp\left( -\frac{(h_i - h_0)^2}{2 \Delta^2} \right).
    \label{eq:signal_term}
\end{equation}
where, $\Delta^2 = N^\gamma\kappa^2 (\mu_g^2+\sigma_g^2+a)$. 

Under the approximation of Eq.~\ref{eq:signal_term} the random fields, $\{ h_i \}$, are i.i.d. random variables with a variance scaling as $N^\gamma$. This variability in the random fields induces heterogeneity in the responses of the neurons in the processing layer, $\{ s_i \}$. In the weak to moderate coarse-tuning regimes, $-1 \le \gamma <0$ this heterogeneity vanishes as the population size, $N$ grows. We therefore focus on the  strong coarse-tuning regime, $\gamma=0$, where heterogeneity in both the random fields and the neuronal responses remains $\mathcal{O} (N^0)$. Note that in the above approximation, the trial-to-trial fluctuations of responses of neurons in the processing layer, $\{ s_i \}$, are uncorrelated. Hence, we  set $\gamma=0$. 

To estimate the stimulus from the neural responses, $\{ s_i \}$, we need to define a readout. A natural choice is a linear readout: $\sum_k c_k s_k $, where $c_k$ are the optimal weights for the processing layer. Consistency requires us to assume strong coarse-tuning in the readout weights from the processing layer as well. Consequently, adding feedforward layers cannot overcome the limiting effect of strong coarse-tuning. Below we consider the addition of recurrent connections to the processing layer.

\subsection{Recurrent Neural Network}
\label{sec:RNN}

Can recurrent connectivity overcome the quenched noise in the case of strong coarse-tuning? To address this question we incorporate recurrent interactions between neurons in the processing layer. Thus, we consider the network to comprise three layers: (i) a representation layer, $ \{ r_i \}_{i=1}^N$, (ii) a single intermediate processing layer with neurons, $ \{ s_i \}_{i=1}^N$, which are recurrently connected, and (iii) a decision unit denoted by $\hat{m}$, Fig.\ref{fig:fig4b}. 

The representation layer $\mathbf{r} = \{ r_i \}_{i=1}^N$ encodes the binary stimulus as previously, and has feedforward connections to the processing layer. Each processing neuron receives an i.i.d. quenched random local field $\{ h_i \}$, distributed according to Equation \eqref{eq:signal_term}, which reflects the input from $\mathbf{r}$. Crucially, unlike the purely feedforward architecture, the processing neurons could now interact via recurrent connections.

Each neuron in the processing layer is modeled as an Ising spin, $s_i\in\{\pm1\}$ for $i \in \{1 \ldots N \}$. The network evolves according to Glauber dynamics at temperature $\frac{1}{\beta}$, governed by the Hamiltonian:
\begin{equation}
    \label{eq:Hamiltonian}
    \mathcal H = -\sum_{i,j}J_{i,j}s_is_j-\sum_i h_is_i.
\end{equation}
The recurrent connections \(\{ J_{i,j} \}\), are taken to be i.i.d. quenched Gaussian random variables with:
\begin{equation}
    \Pr(J_{i,j}) = \sqrt\frac{N}{2\pi J^2} 
    \exp\left( -\frac{N(J_{i,j} - \frac{J_0}{N})^2}{2 J^2} \right),
    \label{eq:interaction_term}
\end{equation}
where $\frac{J_0}{N}$ is the mean strength of the recurrent connections and $\frac{J}{\sqrt{N}}$ is the standard deviation of the recurrent connections. The parameter $J$ represents strong quenched disorder in the recurrent connections, similar to the strong coarse-tuning of the feedforward connectivity with $\gamma = 0$. Thus, the stochastic response of the Recurrent Neural Network neurons is naturally mapped to the Sherrington-Kirkpatrick (SK) model \cite{sherrington1975solvable} with random fields \cite{soares1994effects, bray1980some}. For simplicity, we assume that the trial-to-trial (thermal) fluctuations are more pronounced than the quenched noise, $T>J$. In this regime, one of the central order parameters  characterizing the system is its magnetization (per spin), $m = \frac{1}{N} \sum_i s_i$. This system has been studied extensively, and in our regime of interest its phase diagram (for $h_0 = 0$) is characterized by an ordered (ferromagnetic), $m>0$, and unordered (paramagnetic), $m=0$, phase, see \cite{soares1994effects,de1978stability,schneider1977random,bray1980some} and appendix \ref{app:cavity_TAP_SK}.   

When the recurrent connectivity exhibits a ferromagnetic bias, $ J_0 > 0$, the Recurrent Neural Network amplifies the signal in the structured component, $h_0$, of the input fields. Figs.\ \ref{fig:Heatmap_Ru_delta_half} \& \ref{fig:Heatmap_Ru_delta_one} illustrate this susceptibility, which quantifies the change in the Recurrent Neural Network response, $m$, to small changes in the input signal, $\chi=\frac{dm}{dh_0}$. As shown, the susceptibility is maximal near the phase transition (marked by the black line). Comparing  Figs.\ \ref{fig:Heatmap_Ru_delta_half} \& \ref{fig:Heatmap_Ru_delta_one}, the main effect of the quenched disorder in the random fields, $\Delta$, is that to obtain high signal amplification for larger values of $\Delta$ a stronger ferromagnetic bias, $J_0$, is required. In the ordered phase, the structured component of the input fields acts as an ordering field that chooses the direction of the spontaneous magnetization, Figs.\ \ref{fig:Heatmap_m_delta_half} \& \ref{fig:Heatmap_m_delta_one}. As in the case of susceptibility, larger values of $\Delta$ require a stronger ferromagnetic bias, $J_0$, to obtain the same level of magnetization.

In the regions of the phase diagram in which the amplification is large, most of the signal resides in a two dimensional subspace spanned by the input fields, $h$, and the uniform direction, Figs.\ \ref{fig:Heatmap_abs_cu_uh}  \& \ref{fig:Heatmap_abs_ch_uh}. This takes us back to our initial problem: extracting the signal requires some degree of tuning of the readout weights. 
\newpage
\section{Discussion}

Here we investigated how coarse-tuning constrains the accuracy of population codes. We focused on the scaling of the signal-to-noise ratio ($SNR$) with population size for two readout algorithms: a na\"{i}ve readout and an optimal readout. We found that the $SNR$ of the na\"{i}ve readout was largely insensitive to coarse-tuning, since its performance is already limited by noise correlations in the neural responses. In contrast, the $SNR$ of the optimal readout exhibited distinct scaling regimes: it grew linearly with population size for weak coarse-tuning, crossed over to sublinear growth at intermediate coarse-tuning, and ultimately saturated for strong coarse-tuning.

Consequently, in the strong coarse-tuning regime, the $SNR$ of both readout algorithms converged to finite limits, albeit for different underlying reasons, for large $N$. The asymptotic $SNR$ of the optimal readout was governed by the magnitude of tuning fluctuations, quantified by $\kappa$, whereas the asymptotic performance of the na\"{i}ve readout was dominated by noise correlations in the population activity. Finally, we showed that this fundamental limitation cannot be alleviated by adding further processing layers or by introducing recurrent connectivity.

How can our theory account for the high levels of accuracy observed in neural systems? One possibility is that the central nervous system operates in the weak to moderate coarse-tuning regimes, in which arbitrarily high accuracy can, in principle, be attained by an optimal readout given a sufficiently large population. However, we showed that in these regimes, the intrinsic heterogeneity of neuronal responses vanishes. This conflicts with extensive experimental evidence of persistent response heterogeneity \cite{lim2018development,ringach2002orientation}, which  cast doubt on the hypothesis that neural circuits operate predominantly in the weak to moderate coarse-tuning regimes.

Alternatively, high decoding accuracy may be achieved in the strong coarse-tuning regime provided that the asymptotic $SNR$ is sufficiently large. In this regime, the saturation of the $SNR$ implies that pooling information beyond an effective population size, $N_{ \mathrm{eff}}$, yields little additional benefits. Hence, the readout is robust to neuronal loss as long as the remaining population exceeds $N_{ \mathrm{eff}}$ neurons.

Synaptic weights are inherently volatile, even in the absence of neuronal activity \cite{ziv2018synaptic,yasumatsu2008principles,hazan2020activity}. As a result, noise in synaptic strengths is expected to accumulate over time. A growing body of experimental evidence indicates that the response statistics of the representation layer can also evolve over time, a phenomenon commonly referred to as \emph{representational drift} \cite{ziv2013long,driscoll2017dynamic,deitch2021representational,khatib2023active,geva2023time}. Together, these sources of variability are expected to degrade the performance of any fixed readout over time. Ongoing learning can partially counteract this degradation by continuously adjusting synaptic weights toward their target values. The interplay between stochastic synaptic fluctuations and corrective learning dynamics should thus produce a steady-state distribution of synaptic weights centered around their fine-tuned values. In the present work, we did not explicitly model the stochastic dynamics of synaptic plasticity; instead, quenched disorder was used to represent this steady-state distribution. A detailed investigation of the coupled dynamics of synaptic volatility and learning is beyond the scope of this study and is left for future work. Nevertheless, we believe that the present framework provides a foundational step in this direction.

With respect to learning, it is important to note that the na\"{i}ve and optimal readouts make markedly different assumptions concerning the capabilities of the central nervous system. Learning the optimal readout generally requires a supervised learning mechanism, whereas the na\"{i}ve readout can be acquired through comparatively simple forms of plasticity, such as homeostatic regulation. In particular, in our model, any homeostatic plasticity rule that induces independent fluctuations of synaptic weights around a common nonzero mean is sufficient to implement the na\"{i}ve readout.

It is often assumed that the central nervous system operates near optimality \cite{friston2010free}. However, because the performance of the na\"{i}ve readout is comparable to that of the optimal readout and because the na\"{i}ve readout can be implemented through substantially simpler learning mechanisms, the na\"{i}ve readout remains a viable alternative. This reasoning is supported by two other potentially counterintuitive considerations.

The first is the invariant-manifold hypothesis which posits that the brain maintains functional stability despite ongoing synaptic volatility because the set of synaptic configurations that support a given computation forms a manifold (or low-dimensional subspace). Synaptic fluctuations are therefore substantial along this manifold, because they are strongly suppressed in directions orthogonal to it, thereby preserving function \cite{susman2019stable,shimizu2021computational,rule2019causes,rule2020stable,dubreuil2022role,driscoll2022representational,sherf2025synaptic,socolovsky2025functional,driscoll2017dynamic}. 
In the context of our work, functionality was measured as the ability to extract the signal from the neuronal responses and estimate the stimulus. The signal was embedded by the response selectivity vector, $\mathbf{g}$ defined in Eq.\ \ref{eq:response_ selectivity}. This vector can be decomposed into a fixed uniform component and zero-mean quenched fluctuations. The uniform direction, corresponding to the na\"{i}ve readout, therefore spans the invariant manifold.

The second consideration arises from comparisons between theoretical readout performance and psycho-physical measurements. Previous work has suggested that the na\"{i}ve readout may be more consistent with empirical observations than the optimal readout \cite{shamir2014emerging}. Mendelson and Shamir \cite{mendels2018relating} estimated the asymptotic accuracy of a na\"{i}ve decoder (specifically, the population vector for angular estimation) and found performance levels comparable to those observed in psycho-physical experiments  \cite{andrews1967perception,dick1989visual}. Taken together, these observations raise the possibility that na\"{i}ve readout mechanisms may play a role in neural decoding.
\begin{acknowledgments}
This work was supported by the Israel Science Foundation (ISF) under Grant No.~624/22 and Grant No.~824/21.
The authors declare no competing interests.
\end{acknowledgments}
\section*{Data Availability Statement}
The code and simulation outputs supporting the findings of this study are openly available on Zenodo at DOI: 10.5281/zenodo.18664876. \cite{hendler2026_zenodo}

\appendix


\section{$SNR^2$ of the fine-tuned na$\ddot{\i}$ve decoder }
\label{app:Naive_SNR2}
Substituting the na$\ddot{\i}$ve choice of weights, $\boldsymbol{w}=\frac{\boldsymbol{1}}{N}$, into Eq.~\ref{eq:SNR_def}, yields
\begin{align}
    signal &= \boldsymbol{w}^{\top}\boldsymbol{g} = \frac{1}{N}\sum_{i=1}^{N} g_i,
    \\
\label{eq:naive_noise}
        noise^2 &= 2\boldsymbol{w}^{\top}\mathbf{C}\boldsymbol{w}=\frac{2}{N^2}\sum_{i,j}C_{ij}=\frac{2aN[1+(N-1)c]}{N^2},
\end{align}
where we use in Eq.\ \ref{eq:naive_noise} the correlation matrix of Eq.~\ref{eq:covariance_matrix}. The $signal$ is a random variable because it  depends on the specific realization of the neuronal response selectivity, $\{g_i\}$. Since the $g_i$'s are i.i.d.\ random variables the $signal$ is a self-averaging quantity with $signal \sim \mathcal{N} \Big(\mu_g + \frac{\sigma_g^2}{N}, \ \frac{\sigma_g^2}{N}\Big)$; thus, in the limit of large $N$ we can ignore the quenched variability of the $signal$ and take it to be deterministic with  $signal = \mu_g$. From Eq.~\ref{eq:covariance_matrix}, the $noise^2$ has no quenched fluctuations and is $\mathcal{O}(N^0)$. Thus, the $SNR^2$ of the na\"{i}ve readout is 
\begin{equation}
     \ll SNR^2(\mathbf{w}^{\text{na\"ive}})\gg
    \approx \frac{N\mu_g^2}{2a[1+(N-1)c]}
\end{equation}


\section{$SNR^2$ of fine-tuned Optimal decoder }
\label{app:Optimal_SNR2}
Substituting the optimal weights, $\boldsymbol{w}^{\mathrm{opt}} =\mathbf{C}^{-1}\boldsymbol{g}$, into Eq.~\ref{eq:SNR_def}, yields
\begin{equation}
    SNR^2(\mathbf{w}^{\mathrm{opt}}) = \frac{\boldsymbol{g}^{\top}\mathbf{C}^{-1}\boldsymbol{g}}{2}.
    \label{eq:SNR2_opt_def}
\end{equation}
For our choice of correlation structure, Eq.\ \ref{eq:naive_noise}, its inverse is given by
\begin{equation}
    \mathbf{C}^{-1}=\frac{1}{a(1-c)}\mathbf{I}-\frac{c}{a(1-c)\big(1-c+Nc\big)}\boldsymbol{1},
\end{equation}
where $\mathbf{I}$ is the identity matrix and $\boldsymbol{1}$ is an all ones matrix.
Thus, the $SNR^2$ is a quenched random variable:
\begin{equation}
    SNR^2(\mathbf{w}^{\mathrm{opt}}) 
    =\frac{1}{2a(1-c)}\sum_{i} g_i^2
    -\frac{c\left(\sum_{i} g_i\right)^2}{a(1-c)\big(1-c+Nc\big)}, 
    \label{SNR2_opt}
\end{equation}
with mean
\begin{equation}
    \ll SNR^2(w^{\mathrm{opt}})\gg 
    = 
    \frac{    \Bigl(
    \bigl(1+(N-2)c\bigr)\sigma_g^2 +
    \bigl(1-c\bigr)\mu_g^2
    \Bigr)}{2a(1-c)\Bigl(\frac{1+c(N-1)}{N}\Bigr)}.
\end{equation}
In the limit of large $N$, $SNR^2(w^{opt}) \approx A_1 + A_2$, where $A_1=\frac{N}{2a(1-c)}\bar{g^2}$, $A_2=\frac{N}{2a(1-c)}\bar{g}^2$, and $\bar{x} \equiv \frac{1}{N} \sum_{i=1}^N x_i$ denotes spatial averaging of $x$. As both $A_1$ and $A_2$ are self averaging, so is $SNR^2(w^{opt})$.


\section{The effect of coarse-tuning on the na$\ddot{\i}$ve decoder}
\label{app:Naive_Decoder_under_Coarse_Tuning}
The $signal$, $s$, of the coarse-tuned na$\ddot{\i}$ve decoder, $ s = \bar{g} + \boldsymbol{\xi}^{\top}\boldsymbol{g}$, 
is a random variable with quenched mean of 
\begin{equation}
    \ll s \gg = \mu_g.
    \label{eq:signal_mean}
\end{equation}
Denoting the quenched fluctuations of the $signal$ by $\Delta s=s-\ll s \gg$, the quenched variance of the $signal$, $\ll (\Delta s)^2 \gg$ is given by
\begin{equation}
    \ll (\Delta s)^2 \gg
    =\frac{\sigma_g^2}{N}+\kappa^2 N^{\gamma} (\mu_g^2+\sigma_g^2).
    \label{eq:signal_var}
\end{equation}

The squared $noise$, $n^2$, of the na\"{i}ve decoder is now a random variable
\begin{equation}
    n^2=2\boldsymbol{w}^{\top}\mathbf{C}\boldsymbol{w}
    =\frac{2}{N^2}\sum_{i,j} C_{ij} + 2I_1 + 2I_2, 
    \label{eq:n2_def}
\end{equation}
 with  
\begin{equation}
    I_1 = \boldsymbol{\xi}^{\top}\mathbf{C}  \boldsymbol{\xi},
    \qquad
    I_2 = \frac{2}{N}\sum_{i,j} C_{ij}\,\xi_j.
    \label{eq:I1I2_defs}
\end{equation}
The quenched mean of the squared $noise$ is given by
\begin{equation}
    \ll n^2 \gg=\frac{2a\,[1+(N-1)c]}{N}+2a\,\kappa^2 N^{\gamma}.
    \label{eq:n2_mean}
\end{equation}
The quenched fluctuations of the squared $noise$, $\Delta n^2 = n^2 - \ll n^2 \gg$, can be written as the sum of two terms: $\Delta n^2 = \Delta I_1 + I_2$, where 
\begin{align}
\nonumber
    \ll (\Delta I_1)^2 \gg
    &= N a^2\,\ll (\xi_i^2-\ll \xi_i^2 \gg)^2 \gg
    \\
    &+2N(N-1)(ac)^2\,\ll \sum_{i\neq j}\sum_{i'\neq j'} \xi_i\xi_j\xi_{i'}\xi_{j'} \gg 
    \\
    &= 2 a^2 \kappa^4 N^{2\gamma-1} \left(  1
    +(N-1) c^2  \right),
    \label{eq:I1_var}
    \\
    \ll (\Delta I_2)^2 \gg
    &= \frac{4}{N^2}\sum_{i,j,i',j'} C_{ij} C_{i'j'}\,\ll \xi_j \xi_{j'} \gg
    \approx 4 (ac)^2 \kappa^2 N^{\gamma}.
    \label{eq:I2_var}
\end{align}

Thus, the quenched variance of $n^2$ is
\begin{equation}
    \ll (\Delta n^2)^2 \gg
    \approx
     8  a^2 \kappa^4 N^{2\gamma-1} (1 + c^2(N-1))
    +16 (a \kappa c)^2 N^{\gamma} .
    \label{eq:n2_var_final}
\end{equation}


\section{Optimal readout under coarse-tuning}
\label{app:Optimal_Decoder_under_Coarse_Tuning}
The quenched mean of the $signal$ of the coarse-tuned optimal readout is
\begin{equation}
    \ll s \gg = \sigma_g.
    \label{eq:opt_s_mean}
\end{equation}
The quenched fluctuations of the $signal$ are given by: 
\begin{equation}
    \Delta s \equiv s - \ll s \gg = \frac{\boldsymbol{g}^{\top}(\boldsymbol{g} - \boldsymbol{\bar{g}})}{N\sigma_g}-\sigma_g + \sum_i\xi_i g_i.
    \label{eq:opt_Delta_s}
\end{equation}
where, assuming large $N$, we approximated $ \mathbf{w}^{\mathrm{opt}} \propto \mathbf{g} - \bar{ \mathbf{g}} $.   

The quenched variance reads:
\begin{equation}
\begin{split}
\ll (\Delta s)^2 \gg
&= \frac{1}{N^2 \sigma_g^2}\,
   \ll \big( (\boldsymbol{g}-\boldsymbol{\bar{g}})^{\top}\boldsymbol{g}
   - N\sigma_g^2 \big)^2 \gg \\
&\quad + \ll \Big( \sum_{i=1}^{N} \xi_i g_i \Big)^2 \gg .
\end{split}
\label{eq:opt_var_split}
\end{equation}
The contribution of the first term on the right hand side of Eq.\ \ref{eq:opt_var_split} can be neglected for large $N$. Thus, the quenched variance of the $signal$ is
\begin{equation}
    \ll (\Delta s)^2\gg
    =\kappa^2 N^{\gamma}(\mu_g^2+\sigma_g^2).
    \label{eq:opt_var_s}
\end{equation}

The squared $noise$ is given by
\begin{align}
    n^2 &= 2\boldsymbol{w}^{\top}\mathbf{C} \boldsymbol{w} = I_3 + I_4 + I_5 
    \\
    I_3 &= \frac{2}{N^2\sigma_g^2}(\boldsymbol{g}-\boldsymbol{\bar{g}})^{\top}\mathbf{C} (\boldsymbol{g}-\boldsymbol{\bar{g}})
    \\
    I_4 &= 2 \boldsymbol{\xi}^{\top} \mathbf{C} \boldsymbol{\xi}
    \\
    I_5 &= \frac{4}{N\sigma_g}(\boldsymbol{g}-\boldsymbol{\bar{g}})^{\top}\mathbf{C} \boldsymbol{\xi}
    \label{eq:opt_n2_def}
\end{align}
and its quenched mean is
\begin{equation}
\begin{split}
\ll n^2 \gg
&= \frac{2}{N^2\sigma_g^2}
   \ll (\boldsymbol{g}-\boldsymbol{\bar g})^{\top}
   \mathbf{C}
   (\boldsymbol{g}-\boldsymbol{\bar g}) \gg \\
&\quad + 2 \ll \boldsymbol{\xi}^{\top}\mathbf{C}\boldsymbol{\xi} \gg \\
&= \frac{2a(1-c)}{N} + 2a \kappa^2 N^{\gamma}.
\end{split}
\label{eq:opt_n2_mean}
\end{equation}
To compute the quenched variance of $n^2$, note that $I_3$ is the $noise$ term of the fine-tuned optimal readout; hence, is self-averaging (see \ref{app:Optimal_SNR2}). Now
\begin{equation}
    \ll (\Delta I_4)^2 \gg
    =8 a^2 \kappa^4 N^{2\gamma-1} \left( 1 + c^2(N-1) \right),
    \label{eq:opt_I1_var}
\end{equation}
and

\begin{equation}
\begin{split}
\ll (\Delta I_5)^2 \gg
&= \frac{16 a^2(1-c)^2}{N^2\sigma_g^2}\,
   \ll \big( \boldsymbol{\xi}^{\top}
   (\boldsymbol{g}-\boldsymbol{\bar g}) \big)^2 \gg \\
&= 16 a^2(1-c)^2 \kappa^2 N^{\gamma-2}.
\end{split}
\label{eq:opt_I5_var}
\end{equation}

where we use in Eq.\ \ref{eq:opt_I5_var} the fact the $(\mathbf{g} - \mathbf{ \bar{g}}) $ is an eigenvector of $ \mathbf{C}$ with eigenvalue of $a(1-c)$. Thus, the quenched variance of \(n^2\) is given by
\begin{equation}
\begin{aligned}    
    \ll (\Delta n^2)^2 \gg
    \approx
    &8 a^2 \kappa^4 N^{2\gamma-1} \left( 1 + c^2(N-1) \right) \\
    &+16 (a \kappa)^2 (1-c)^2 N^{\gamma-2}.
    \label{eq:opt_n2_var_final}
\end{aligned}
\end{equation}


\section{Feedforward neural network}
\label{app:ffnn}

We consider a processing layer of \(N\) neurons with local fields
\begin{equation}
    h_i = \sum_{j=1}^{N} \tilde{w}_{ij} (\mathbf{r}^t - \mathbf{r}^d)
    \label{eq:ffnn_h_def}
\end{equation}
where, 
\begin{equation}
\begin{split}
\tilde{w}_{ij}
&= \frac{w_{ij}}{\sqrt{N}\,\|\mathbf{w}\|} + \xi_{ij}, \\
\ll \xi_{ij} \gg
&= 0,\qquad
\ll \xi_{ij} \xi_{i'j'} \gg
= \kappa^2 N^{\gamma-1}\delta_{ii'}\delta_{jj'} .
\end{split}
\label{eq:weights_model}
\end{equation}
The mean of the local field is
\begin{equation}
    [h_i] 
    = \frac{ \ll\mathbf{w}^\top \mathbf{g} \gg}{\sqrt{N} \|\mathbf{w}\|} = \frac{ \ll(\mathbf{g}-\bar{\mathbf{g}})^{\top} \mathbf{g} \gg}{\sqrt{N} \|\mathbf{w}\|}
    \equiv h_0,
    \label{eq:hi_mean}
\end{equation}
where $[\cdot]$ denotes averaging with respect to both the trial-to-trial variability and the quenched disorder. The covariance of the local fields is given by
\begin{equation}
\begin{split}
\mathrm{cov}(h_i,h_k)
&= [h_i h_k] - h_0^2 \\
&= \big[ (h_i - \langle h_i\rangle)
         (h_k - \langle h_k\rangle) \big] \\
&+ \big[ (\langle h_i\rangle - [h_i])
               (\langle h_k\rangle - [h_k]) \big] .
\end{split}
\label{eq:hi_cov}
\end{equation}
The first term on the right hand side of Eq.\ \ref{eq:hi_cov} yields
\begin{equation}
\begin{aligned}
    [ (h_i - \langle h_i \rangle)(h_k - \langle h_k \rangle) ]
    &= [ \mathbf{w} C \mathbf{w} ] + a N^\gamma \kappa^2 \delta_{ik} \\
    &\approx \frac{a(1-c)}{N} + a N^\gamma \kappa^2 \delta_{ik},
\end{aligned}
\end{equation}
where for large $N$ we approximated $
    \mathbf{w} C \mathbf{w} \approx \frac{1}{N^2\sigma_g^2} (\mathbf{g}-\bar{\mathbf{g}})^\top C (\mathbf{g}-\bar{\mathbf{g}})
    = \frac{a(1-c)}{N}$.
The second term yields
\begin{equation}
    [(\langle h_i \rangle - [h_i])(\langle h_k \rangle - [h_k])]
    = N^\gamma\kappa^2(\mu_g^2+\sigma_g^2) \delta_{ik}.
\end{equation}
Thus, we obtain
\begin{equation}
    \mathrm{cov}(h_i,h_k) =
    \begin{cases}
        \dfrac{a(1-c)}{N}+N^\gamma\kappa^2(a + \mu_g^2+\sigma_g^2), & i=k,\\[6pt]
        \dfrac{a(1-c)}{N}, & i\ne k.
    \end{cases}
\end{equation}


\section{Sherrington-Kirkpatrick model}
\label{app:rfim_crit}
The Recurrent Neural Network model of section \ref{sec:RNN} naturally maps to the Sherrington-Kirkpatrick model with random fields. This model has been extensively studied; see e.g., \cite{edwards1975theory, sherrington1975solvable, soares1994effects, bray1980some}. For completeness we summarize the central results that are relevant to the current study.

We consider a system of $N$ Ising spins, $ \{ s_i \}_{i=1}^N$, governed by the Hamiltonian in Eq.\ \ref{eq:Hamiltonian} at thermal equilibrium with inverse temperature $\beta$. The fields, $\{ h_i \}$, are i.i.d.\ quenched Gaussian random variables, with the probability density 
\begin{equation}
    \Pr(h_i) 
    = \frac{1}{ \sqrt{2\pi} \Delta} 
    \exp\left( -\frac{(h_i - h_0)^2}{2 \Delta^2} \right).
    \label{eq:signal_term_appendix}
\end{equation}
The recurrent connections, \(\{ J_{i,j} \}\), are i.i.d.\ quenched Gaussian random variables with 
\begin{equation}
    \Pr(J_{i,j}) = \sqrt\frac{N}{2\pi J^2} 
    \exp\left( -\frac{N(J_{i,j} - \frac{J_0}{N})^2}{2 J^2} \right).
    \label{eq:interaction_term_appendix}
\end{equation}

To appreciate the effect of recurrent connectivity it is insightful to first consider the case of $J=0$. At thermal equilibrium the local magnetization, $m_i = \langle s_i \rangle$, is given by: 
\begin{equation}
\label{eq:local_mag1}
    m_i=\tanh\big(\beta(J_0 m + h_i)\big),
\end{equation} 
where $m$ is the mean magnetization, $m= \frac{1}{N} \sum_i m_i$, which serves as an order parameter of the system. Thus, the self-consistent equation for $m$ is  
\begin{equation}
\label{eq:MforJequal0}
    m=\int \mathcal{D}z \tanh\big(\beta(J_0m+h_0 + \Delta z)\big),
\end{equation}
where $\mathcal{D}z\equiv \frac{dz}{\sqrt{2\pi}} \exp{\frac{-z^2}{2}}$ is the standard Gaussian measure. In the absence of external fields, $h_0 = 0$ and $\Delta = 0$, the paramagnetic state, $m=0$, is always a solution of Eq.\ \eqref{eq:MforJequal0}.
When the recurrent connectivity, $J_0$, is stronger than some critical value, $J_0^{\text{crit}} = \frac{1}{\beta}$, the paramagnetic solution loses its stability and the system settles into one of the two solutions of the ferromagnetic phase. In this case, even a small ordering field, $ | h_0 | \ll 1$, will choose  the sign of the ferromagnetic solution. Thus, in this case, the recurrent connectivity offers amplification of its input.  

To understand the effect of random fields, $\Delta >0$, we first examine the case $J=0$ and $h_0 = 0$. In this case, due to the symmetry of Eq.\ \eqref{eq:MforJequal0}, a solution with $m=0$ always exists. However, since $\Delta >0$ the local magnetization cannot be zero. Thus, the system is characterized by an additional order parameter:
\begin{equation}
q = \int \mathcal{D}z \tanh^{2}\big(\beta(J_0 m + h_0 + \Delta z)\big),
\label{eq:q_order_parameter}
\end{equation}
the Edward Anderson order parameter \cite{edwards1975theory}. 

To find the the ferromagnetic transition (see \cite{schneider1977random}), one sets the mean of the external field to zero, $h_0=0$, and, assuming a second order phase transition, expands Eq.\ \ref{eq:MforJequal0} in small $m$. 
Using
\begin{equation}
\begin{split}
    \tanh(x+y) & =\tanh x \\
    & +y(1-\tanh^{2}x) \\
    & -y^{2}\tanh x(1-\tanh^{2}x) \\
    & +\frac{y^{3}}{3}\left((1-\tanh^2 x)(3\tanh^2 x-1)\right) \\
    & -\frac{y^{4}}{3}\left(\tanh x\big(1-\tanh^2 x\big)\big(3\tanh^2 x-2\big)
    \right) \\
    & +y^{5} (-\tanh^{6}x +2\tanh^{4}x - \frac{17}{15}\tanh^{2}x \\
    &+ \frac{2}{15})+ O(y^{6}),
\end{split}
\end{equation}
one obtains
\begin{equation}
\begin{split}
    m &= (\beta J_0 m)\int \mathcal{D}z\,\big(1-\tanh^{2}[\beta\Delta z]\big) \\
      & + \frac{(\beta J_0 m)^3}{3} \int \mathcal{D}z\,\Big(1-\tanh^2[\beta\Delta z]\Big)\Big(3\tanh^2[\beta\Delta z]-1\Big) \\
      & + (\beta J_0 m)^5 \int \mathcal{D}z\,\Big(-\tanh^{6}[\beta\Delta z] + 2\,\tanh^{4}[\beta\Delta z] \\
      &- \tfrac{17}{15} \tanh^{2}[\beta\Delta z] + \tfrac{2}{15}\Big) \\
      & + O(m^{7}).
\end{split}
\end{equation}
Equivalently, denoting \(q_{2k}\equiv\int \mathcal{D}z\tanh^{2k}(\beta\Delta z)\),
\begin{equation}
\begin{split}
    m &= \beta J_0(1-q_2)m
    + (\beta J_0)^3\left(-q_4 + \tfrac{4}{3}q_2 - \tfrac{1}{3}\right)m^3\\
    &+ (\beta J_0)^5\left(-q_6 + 2q_4 - \tfrac{17}{15}q_2 + \tfrac{2}{15}\right)m^5
    + O(m^{7}).
\end{split}
\label{eq:mvsq}
\end{equation}
A second-order phase transition to the ferromagnetic phase occurs when the coefficient of the linear term in $m$ in Eq.~\ref{eq:mvsq} changes sign
\begin{equation}
    \beta_c J_{0,c}\big(1 - q(\beta_c, J_{0,c})\big) = 1.
\end{equation}
Note that under the assumption of Gaussian random fields, the coefficient of the cubic term in $m$ in Eq.~\ref{eq:mvsq} is always positive; see \cite{aharony1978tricritical}. 
Accordingly, a transition to the ferromagnetic phase can always be induced for sufficiently large $J_0$. However, in the presence of random fields ($\Delta>0$), the transition is shifted to larger values of the recurrent coupling $J_0$. Thus, fluctuations in the local fields           ($\Delta>0$) attenuate the amplification of the ordering field $h_0$ by the recurrent interactions.

In the general case of strong quenched disorder in the recurrent connections, $J>0$, one can derive equations for the local magnetization that generalizes Eq.~\eqref{eq:local_mag1} (see Appendix \ref{app:cavity_TAP_SK}):
\begin{equation}
    m_i=
    \tanh\Bigl(
    \beta\sum_{j(\neq i)}J_{ij}m_j
    -\beta^{2}J^{2}(1-q)m_i + \beta h_i
    \Bigr).
    \label{eq:TAP_SK}
\end{equation}
Within our chosen parameter regime, $\beta J < 1$, the two order parameters $m$ and $q$ suffice to characterize the system \cite{edwards1975theory}. Averaging Eq.~\eqref{eq:TAP_SK} over space and replacing the spatial average with averaging over the quenched disorder yields the self-consistent equations for the order parameters:
\begin{align}
    m = \int \mathcal{D}z \tanh\Big(\beta (J_0 m + h_0 + \sqrt{\Delta^{2} + J^{2} q} z)\Big),
    \label{eq:m_full}\\
    q = \int \mathcal{D}z \tanh^{2}\bigl(\beta \bigl(J_0 m + h_0 + \sqrt{\Delta^{2} + J^{2} q} z\bigr)\bigr).
    \label{eq:q_full}
\end{align} 
Similar to random fields ($\Delta>0$), the coupling of quenched disorder in the recurrent connections ($J>0$) with nonzero $q$, suppresses amplification of the ordering field, $h_0$. As a result, the transition to the ferromagnetic phase is shifted to larger values of the ferromagnetic bias $J_0$. Nevertheless, for sufficiently large $J_0$, the recurrent interactions still provide amplification of the signal, $h_0$. 

We further map the Almeida-Thouless (AT) line \cite{de1978stability}, writing
\begin{equation}
    \mathcal{A} = (\beta J)^2 \int \mathcal{D}z \mathrm{sech}^{4}\big(\beta\big(J_0 m + h_0 + \sqrt{\Delta^{2} + J^{2} q}\, z\big)\big).
\end{equation}
For \(\mathcal{A}<1\), the two order parameters in \eqref{eq:m_full} and \eqref{eq:q_full} fully characterize the system. The AT boundary is given by \(\mathcal{A}=1\); beyond it (\(\mathcal{A}>1\)) the solution above is insufficient and additional techniques need to be  utilized to characterize the system such as replica symmetry breaking \cite{bray1979replica, blandin1980mean, parisi1979infinite}.


\section{TAP equation}
\label{app:cavity_TAP_SK}

Here we derive TAP equations \cite{thouless1977solution} for the Sherrington-Kirkpatrick model with local random fields using the cavity method \cite{mezard1987spin}. The derivation presented below extends the method of Ref. \cite{shamir2000thouless} to include random local fields. The extension is straightforward and is presented here for completeness. The derivation proceeds in three steps.


\subsection*{Step 1: adding a spin}
We first extend the $N$-spin system by adding a single spin, $s_0$, at site $0$, which interacts with the existing spins via couplings $\{J_{0j}\}_{j=1}^N$ drawn from the distribution in Eq.~\ref{eq:interaction_term_appendix}. The Hamiltonian of the extended $(N+1)$-spin system is given by
\begin{equation}
    \mathcal H^{(N+1)} = \mathcal H^{(N)} - h_0 s_0 ,
    \label{eq:app_HN1}
\end{equation}
\begin{equation}
    h_0 = \sum_{j=1}^{N} J_{0j}\,s_j +h_0^{ext}.
    \label{eq:app_h0_def}
\end{equation}
where $h_0$ is the local field at site 0.
The states of the extended system are distributed according to the Gibbs distribution with Hamiltonian $\mathcal H^{(N+1)}$:
\begin{equation}
    P_{N+1}\left(\{s_i\}_{i=0}^{N}\right) = \frac{1}{Z_{N+1}} \exp\big[-\beta \mathcal H^{(N+1)}\big],
\label{eq:P_N+1}
\end{equation}
\begin{equation}
    Z_{N+1} = \operatorname{Tr}_{\{s_i\}_{i=0}^{N}} \exp\big[-\beta \mathcal H^{(N+1)}\big].
\end{equation}

From Eq.\ \ref{eq:P_N+1} we obtain the marginal joint distribution of the cavity spin, $s_0$, and the cavity field, $h_0$,
\begin{equation}
\begin{split}
P_{N+1}(h_0,s_0) &= 
\frac{1}{Z_{N+1}} 
\operatorname{Tr}_{\{s_i\}_{i=1}^{N}} 
\Biggl\{ \,
\delta\Bigl(h_0 - h_0^{\rm ext} - \sum_{j=1}^{N} J_{0j} s_j \Bigr) \\
&\qquad \times \exp\big[-\beta \mathcal{H}^{(N+1)}\big] \Biggr\}.
\end{split}
\end{equation}
Using Eq.\ \ref{eq:app_HN1} we can write 
\begin{equation}
    P_{N+1}(h_0,s_0) = \frac{1}{z}\exp\big[\beta h_0 s_0\big]P_{N}(h_0),
\end{equation}
where 
\begin{equation}
\begin{split}
P_{N}(h_0) &= 
\frac{1}{Z_{N}} 
\operatorname{Tr}_{\{s_i\}_{i=1}^{N}} 
\Biggl\{ \,
\delta\Bigl(h_0 - h_0^{\rm ext} - \sum_{j=1}^{N} J_{0j} s_j \Bigr) \\
&\qquad \times \exp\big[-\beta \mathcal{H}^{(N)}\big] \Biggr\},
\end{split}
\end{equation}
and the normalization, $z$, is given by 
\begin{equation}
    z = \frac{Z_{N+1}}{Z_N} = \big\langle 2\cosh(\beta h_0) \big\rangle_N .
\end{equation}
with $\langle \cdot \rangle_N$ denoting thermal averaging with respect to the $N$-spin system.

Averaging $s_0$ and $h_0$ with $P_{N+1}(h_0,s_0)$ gives the standard identities
\begin{equation}
\label{eq:MeanS0}
    \langle s_0\rangle_{N+1} =
    \frac{ \langle \sinh(\beta h_0) \rangle_N }
         { \langle \cosh(\beta h_0) \rangle_N } ,
\end{equation}
\begin{equation}
\label{eq:Mean_h0}
    \langle h_0\rangle_{N+1} =
    \frac{  \langle h_0 \cosh(\beta h_0) \rangle_N }
         { \langle \cosh(\beta h_0) \rangle_N } .
\end{equation}

\subsection*{Step 2: statistics of the cavity field $h_0$}
Next, we characterize the first two moments of the cavity field, $h_0$, in the $N$-spin system:
\begin{align}
    \langle h_0\rangle_N &= \sum_{j=1}^{N} J_{0j}\, \langle s_j\rangle_N ,
\\
    \langle ( \delta h_0 )^2 \rangle_N
    &= \sum_{i,j=1}^{N} J_{0i} J_{0j}\, \langle \delta s_i \,\delta s_j\rangle_N = \sum_{i=1}^{N} J_{0i}^2\, \langle (\delta s_i)^2\rangle_N,
    \label{eq:delta_h_0}
\end{align}
with $\delta X = X - \langle X \rangle_N$. In the last equality of Eq.\ \ref{eq:delta_h_0} we used the fact that $\langle \delta s_i \delta s_j\rangle_N$ is of order $ \mathcal{O}(1/ \sqrt{N})$ and is independent of $J_{0j}$, to approximate the sum over $i$ and $j$ only by the contribution of the $i=j$ terms.  

Defining $q_N = \frac{1}{N} \sum_{i=1}^{N} \langle s_i\rangle_N^2$, and using the self-averaging of the right-hand side of Eq.\ \ref{eq:delta_h_0} we obtain $\langle ( \delta h_0 )^2 \rangle_N = J^2\,(1-q_N)$. Assuming the distribution of the cavity field, $h_0$, in the $N$ spin system is well-approximated by Gaussian distribution as the sum of $N$ independent variables.  Further assuming that $q_N \to q$ as $N\to\infty$, the distribution of the cavity field is given by:
\begin{equation}
    P_N(h_0) =
    \frac{1}{\sqrt{2\pi J^2(1-q)}}\,
    \exp\left(
    -\frac{\big(h_0 - h_0^{\text{ext}}- \langle h_0\rangle_N\big)^2}{2 J^2(1-q)}
    \right).
    \label{eq:distrebution_of_cavity_field}
    \end{equation}

\subsection*{Step 3: obtaining the TAP equation}
Substituting \eqref{eq:distrebution_of_cavity_field} into the Eqs.\ \ref{eq:MeanS0} \& \ref{eq:Mean_h0} yields
\begin{equation}
\label{eq:s0_N+1}
    \langle s_0\rangle_{N+1} = \tanh\big( \beta \langle h_0\rangle_N \big),
\end{equation}
\begin{equation}
\label{eq:h0_N+1}
    \langle h_0\rangle_{N+1} = \langle h_0\rangle_N + \beta J^2(1-q)\langle s_0\rangle_{N+1}.
\end{equation}
Substituting $\langle h_0\rangle_N$ from Eq.\ \ref{eq:h0_N+1} into Eq.\ \ref{eq:s0_N+1} and replacing site 0 with a generic site $i$, one obtains the TAP equations:
\begin{equation}
    m_i=
    \tanh\Bigl(
    \beta\sum_{j(\neq i)}J_{ij}m_j
    -\beta^{2}J^{2}(1-q)m_i + \beta h_i^{\mathrm{ext}}
    \Bigr).
    \label{eq:TAP}
\end{equation}
Note that, in this case $h_i^{\mathrm{ext}}$ also exhibits quenched statistics.

\subsection*{The susceptibility matrix}

The susceptibility matrix is defined by  \(
\chi_{ik} \equiv \partial m_i/\partial h_k
\). The diagonal element of the susceptibility matrix can be obtained using the fluctuation-dissipation theorem, $ \mathbf{\chi} = \beta \mathbf{C}$, where $\mathbf{C}$ is the correlation matrix $C_{ij} = \langle \delta s_i \delta s_j \rangle$. Since $ s_i \in \{ \pm 1 \}$, $C_{ii} = \langle s_i^2 \rangle - \langle s_i \rangle^2 = 1- m_i^2$. Thus, 
\begin{align}
    \chi_{ii} = \beta (1- m_i^2 ) .
\end{align}

For the off-diagonal elements, we 
differentiate Eq.\ \ref{eq:TAP} with respect to the local field at site $k$, yielding
\begin{equation}
    \chi_{ik}  \equiv \partial m_i/\partial h_k 
    = (1-m_i^{2}) \beta 
    \Bigl[
    \delta_{ik}
    +\sum_{j} J_{ij} \chi_{jk}
    -\beta J^{2}(1-q) \chi_{ik}
    \Bigr].
    \label{eq:chi_start}
\end{equation}
Defining the matrix $\mathbf{D}$ to be
\begin{align}
    D_{ij} = \delta_{ij} \beta (1-m_i^2),
\end{align}
we obtain 
\begin{align}
    \chi_{ij}  = \Big( \mathbf{D}  + \beta J^2 (1-q^2) \mathbf{I} - \mathbf{J} \Big)^{-1}_{ij} & & i \neq j .
\end{align}

\subsection*{Figures \ref{fig:fig1}-\ref{fig:fig3}: Neural population readout}

Table~\ref{tab:params_readout} summarizes the parameter values used for the results in figures \ref{fig:fig1}-\ref{fig:fig3}. Parameters marked with an asterisk ($^*$) vary across figure panels and are specified explicitly in the corresponding captions. 

\begin{table}[h!]
\caption{\label{tab:params_readout}Parameters for the neural population readout model (Figs.~\ref{fig:fig1}--\ref{fig:fig3}).}
\begin{ruledtabular}
\begin{tabular}{llc}
Parameter & Description & Value \\ \hline
$N$              & Number of Neurons                              & $^*$ \\
$a$              & Single-neuron variance                       & $12$ \\
$c$              & Pairwise correlation coefficient             & $^*$ \\
$\mu_t,\mu_d$    & Quenched mean response (t,d)  & $12,\;9$ \\
$\mu_g$          & Quenched mean of response selectivity  & $3$ \\
$\sigma_g^{2}$   & Quenched variance of response selectivity & $24$ \\
$\kappa$         & Coarse-tuning magnitude                      & $^*$ \\
$\gamma$         & Coarse-tuning scaling               & $^*$ \\
\end{tabular}
\end{ruledtabular}
\end{table}

We performed 500 realizations of the quenched disorder. In each realization, we sampled $\mu_t^i \sim \mathcal{N}(\mu_t, \sigma_t^2)$ and $\mu_d^i \sim \mathcal{N}(\mu_d, \sigma_d^2)$ for each neuron $i$, yielding the selectivity $g_i = \mu_t^i - \mu_d^i$. Coarse-tuning was drawn as $\xi_i \sim \mathcal{N}(0, \kappa^2 N^{\gamma+1})$. For each realization, we constructed the readout weights for the na\"{i}ve and optimal decoders and computed the $SNR$. Finally we averaged the $SNR$ over all realizations. The simulation results of the $SNR$ (markers) were compared to  the analytical expressions of the $SNR$ (solid lines) derived in the main text. 

\subsection*{Figure \ref{fig:fig4}: Feedforward Neural Network and Recurrent Neural Network simulations}
Table~\ref{tab:params_spin} summarizes the parameter values used for the Feedforward Neural Network and Recurrent Neural Network simulations. Parameters marked with an asterisk ($^*$) vary across figure panels and are specified explicitly in the corresponding captions. The coupling matrix and local fields were constructed according to Eqs.~(\ref{eq:signal_term}) and (\ref{eq:interaction_term}), respectively.

\begin{table}[h!]
\caption{\label{tab:params_spin}Parameters for the Feedforward Neural Network and Recurrent Neural Network simulations (Fig.~\ref{fig:fig4}).}
\begin{ruledtabular}
\begin{tabular}{llc}
Parameter & Description & Value \\ \hline
$N$              & Number of neurons in processing layer                              & $200$ \\
$J$              & Disorder in the recurrent connections                            & $1$ \\
$h_0$            & Signal                                 & $^*$ \\
$\Delta$       & Disorder in the random fields                            & $^*$ \\
$T$            & Temperature                            & $[1.5,\;3.5]$ \\
$J_0$          & Ferromagnetic bias                 & $[0,\;4]$ \\
\end{tabular}
\end{ruledtabular}
\end{table}

The TAP equations, Eq. (\ref{eq:TAP}), were solved using damped fixed-point iteration with mixing parameter $\alpha=0.35$. Convergence was declared when $\max_i|m_i^{(t+1)}-m_i^{(t)}| < 10^{-10}$, with a maximum of 4000 iterations.

The susceptibility matrix $\chi$, Eq. (\ref{eq:chi_start}), was computed from the converged TAP solution. The dominant eigenvector of $\chi$, $\mathbf{v}_{\mathrm{max}}$ was obtained using MATLAB's eigs function with tolerance $10^{-10}$; if this fails to converge, a full eigendecomposition was performed instead.

To analyze the structure of the dominant mode, Fig. \ref{fig:fig4}g \& \ref{fig:fig4}h, we projected the unit-normalized eigenvector $\hat{\mathbf{v}} = \mathbf{v}_{\mathrm{max}}/\|\mathbf{v}_{\mathrm{max}}\|$ onto the subspace spanned by $\hat{\mathbf{u}} = \mathbf{1}/\|\mathbf{1}\|$ (uniform direction) and $\hat{\mathbf{h}} = \mathbf{h}/\|\mathbf{h}\|$ (normalized field). The projection coefficients $(c_u, c_h)$ were obtained by solving
\begin{equation}
\begin{pmatrix} 1 & \hat{\mathbf{u}} \cdot \hat{\mathbf{h}} \\ \hat{\mathbf{u}} \cdot \hat{\mathbf{h}} & 1 \end{pmatrix} \begin{pmatrix} c_u \\ c_h \end{pmatrix} = \begin{pmatrix} \hat{\mathbf{u}} \cdot \hat{\mathbf{v}} \\ \hat{\mathbf{h}} \cdot \hat{\mathbf{v}} \end{pmatrix}.
\end{equation}
The residual $\boldsymbol{\delta} = \hat{\mathbf{v}} - c_u \hat{\mathbf{u}} - c_h \hat{\mathbf{h}}$ is orthogonal to both basis vectors.

The phase transition for $h_0=0$, denoted in Fig. \ref{fig:fig4}c-f by a black line, was computed as follows. For each temperature, we first solved Eq. (\ref{eq:q_full}) for the Edwards-Anderson parameter $q$. The critical $J_0/J$ is then given by
\begin{equation}
\frac{J_0}{J} = \frac{T/J}{\langle \mathrm{sech}^2(\beta J \sqrt{q + (\Delta/J)^2} \, z) \rangle_z},
\end{equation}
where $z \sim \mathcal{N}(0,1)$ and the expectation was evaluated using Gauss-Hermite quadrature with 80 nodes. All heat maps were computed on a $50 \times 50$ grid for a single disorder realization.

\bibliographystyle{apsrev4-2}
\bibliography{FineTuningRefrences}

\end{document}